\documentclass{aa}  
\usepackage{amssymb}
\usepackage{pifont}
\usepackage{graphicx}
\usepackage{xcolor}
\usepackage{amsmath}
\usepackage{pgfplots}
\usepackage{txfonts}
\usepackage[colorlinks=true,allcolors=blue]{hyperref}
\usepackage{natbib}
\usepackage{placeins}
\bibpunct{(}{)}{;}{a}{}{,}

\begin{document} 

\title{Unveiling the trends between dust attenuation and galaxy properties at $z \sim 2-12$ with the James Webb Space Telescope}
   \titlerunning{Trends between dust attenuation and galaxy properties}
   \author{V. Markov\inst{1, 2}, S. Gallerani\inst{2}, A. Pallottini\inst{3, 2}, M. Brada{\v c}\inst{1, 4}, S. Carniani\inst{2},  R. Tripodi\inst{5}, G. Noirot\inst{6}, F. Di Mascia\inst{2}, E. Parlanti\inst{2},  N. Martis\inst{1}  %
     }
   \authorrunning{Markov et al.}
      \institute{Faculty of Mathematics and Physics, University of Ljubljana, Jadranska ulica 19, SI-1000 Ljubljana, Slovenia\\
              \email{vladan.markov@fmf.uni-lj.si}
         \and
   Scuola Normale Superiore, Piazza dei Cavalieri 7, 56126 Pisa, Italy
   \and
    Dipartimento di Fisica ``Enrico Fermi'', Universit\'{a} di Pisa, Largo Bruno Pontecorvo 3, Pisa I-56127, Italy
    \and 
   Department of Physics and Astronomy, University of California Davis, 1 Shields Avenue, Davis, CA 95616, USA
   \and
   INAF - Osservatorio Astronomico di Roma, Via Frascati 33, Monte Porzio Catone, 00078, Italy
   \and
   Space Telescope Science Institute, 3700 San Martin Drive, Baltimore, Maryland 21218, USA
             }
   \date{Received XX XX, 2025; accepted XX XX, 2025}

  \abstract
   {A large variety of dust attenuation and/or extinction curves has been observed in high-redshift galaxies. Some studies investigated their correlations with fundamental galaxy properties, which yielded mixed results. These variations are likely driven by underlying factors such as the intrinsic dust properties, the total dust content, and the spatial distribution of dust relative to stars.
  }
   {We investigate the correlations between the shape of dust attenuation curves, defined by the UV-optical slope ($S$) and the UV bump parameter ($B$), and fundamental galaxy properties. Our goal is to identify the key physical mechanisms that shape the dust attenuation curves through cosmic time in the broader context of galaxy formation and evolution. 
   }
   {We extended the analysis of 173 dusty high-redshift ($z\sim 2-11.5$) galaxies, whose dust attenuation curves were inferred by fitting {\it James Webb Space Telescope} (JWST) data with a modified version of the spectral energy distribution (SED) fitting code \texttt{BAGPIPES}.
   We investigate the trends between the dust attenuation parameters and different galaxy properties as inferred from the SED fitting: $V$-band attenuation ($A_V$), star formation rate (SFR), stellar mass ($M_*$), specific SFR ($\rm{sSFR = SFR}/M_{*}$), mass-weighted stellar age (${\langle a \rangle}_*^{\rm{m}}$), ionization parameter ($\log{U}$), and metallicity ($Z$). For a subset of sources, we additionally explored the trends with oxygen abundance ($12 + \log(\rm{O/H})$), which we derived using the direct $T_e$-based method.} 
   {We report moderate correlations between $S$ and $A_V$, and $B$ and $A_V$. Galaxies characterized by lower (higher) $A_V$ exhibit steeper (flatter) slopes and stronger (weaker) UV bumps. These results agree with radiative transfer (RT) predictions that account for the total dust content and the relative spatial distribution of dust with respect to stars. 
  Additionally, we find that $S$ flattens with decreasing ${\langle a \rangle}_*^{\rm{m}}$ and increasing sSFR. These two trends can be explained if the strong radiation fields associated with young stars (low ${\langle a \rangle}_*^{\rm{m}}$) and/or bursty galaxies (high sSFR) preferentially destroy small dust grains, which would shift the size distribution toward larger grains. 
  Finally, the positive correlation between $B$ and $12 + \log(\rm{O/H})$ that emerged from our analysis might be driven by variations in the intrinsic dust properties with the gas metallicity.}
  {The shape of the dust attenuation curves primarily correlates with four key galaxy properties: 1) With the redshift, which traces variations in the intrinsic and/or reprocessed dust properties, 2) with $A_V$ , which reflects RT effects, 3) with the mass-weighted stellar age or sSFR, which might be driven by the radiation field strength, and 4) with the oxygen abundance, which might be linked to intrinsic dust properties. The overlap between some of these mechanisms makes it difficult to isolate their contributions, however. Further progress requires a combination of observations of a larger galaxy sample, point-like sources, spatially resolved galaxy studies, and theoretical models incorporating dust evolution in cosmological simulations.}
 
   
   \maketitle

\section{Introduction} \label{intro}

Dust attenuation is defined as the wavelength-dependent absorption and scattering of light from galaxies that is caused by interstellar dust along the line of sight (LOS). It depends on the intrinsic dust properties (primarily, on the grain size distribution and composition) and radiative transfer (RT) effects that take the total dust content along the LOS and the complex distribution of dust relative to stars in galaxies into account. In local galaxies, where resolved stars and foreground dust act as point-like sources behind a uniform screen, this effect is instead called dust extinction and is primarily driven by the intrinsic dust properties (see, e.g., \citealp{2001PASP..113.1449C, 2013seg..book..419C, 2020ARA&A..58..529S} for reviews). 

In the local Universe, dust attenuation and extinction laws range from the relatively flat Calzetti attenuation curve (\citealp{1994ApJ...429..582C, 2000ApJ...533..682C}) to the steeper extinction curve of the Small Magellanic Cloud (SMC) (\citealp{2003ApJ...594..279G, 2024ApJ...970...51G}). The amplitude of the UV bump also varies significantly from being absent, as in the SMC, to the prominent UV bump observed in the Milky Way (MW; \citealp{1989ApJ...345..245C}). Furthermore, extinction curves are diverse even within individual galaxies including the MW (\citealp{1990ApJS...72..163F, 2000ApJS..129..147C, 2007ApJ...663..320F}), SMC, and Large Magellanic Cloud (LMC; \citealp{2003ApJ...594..279G, 2017MNRAS.466.4540H, 2024ApJ...970...51G}). The steepness and the amplitude of the UV bump also varies considerably for individual sight lines.
 
At high redshifts ($z > 3$), the attenuation curves of galaxies (\citealp{2022A&A...663A..50B, 2023A&A...679A..12M, 2024NatAs.tmp...20M,  2024arXiv240805273S, 2025MNRAS.539..109F, ormerod2025detection2175aauvbump}) and the extinction curves of quasars (\citealp{2004Natur.431..533M, 2010A&A...523A..85G, 2021MNRAS.506.3946D}), and gamma-ray burst (GRB) afterglows (\citealp{2011A&A...532A..45S, 2018A&A...609A..62B, 2018MNRAS.480..108Z}) deviate from the standard empirical curves of local galaxies. These deviations likely arise from a combination of factors, that is,  
(i) from variations in the intrinsic dust properties (\citealp{2022MNRAS.517.2076M}) driven by the different dust production mechanisms (\citealp{2024NatAs.tmp...20M}), and
(ii) from differences in the amount and relative spatial distribution of dust with respect to stars (\citealp{2018ApJ...869...70N, 2020MNRAS.491.3937T})
because the dust content at high $z$ is generally lower (\citealp{2023MNRAS.522.3986F, 2024MNRAS.531..997C, 2024arXiv241207598N, 2025A&A...694A.215F, 2025ApJ...985L..21M}) and the dust and gas distribution is more complex and clumpier than in stars at intermediate to high redshifts (\citealp{2018MNRAS.477..552B, 2018MNRAS.478.1170C, 2019ApJ...876..130H,  2020A&A...641A..22M, 2022MNRAS.517.5930S}). 

We adopted the parameterization of \cite{2020ARA&A..58..529S} to characterize the shape of the dust attenuation and/or extinction curves with two key features: the overall ultraviolet (UV)-optical slope ($S$), and the characteristic UV bump strength ($B$) at $\sim 2175 \AA$  (\citealp{2020ARA&A..58..529S}). Alternative parameterizations of these curves exist, however, that incorporate varying numbers of parameters (e.g., \citealp{2007ApJ...663..320F, 2008ApJ...685.1046L, 2009A&A...507.1793N, 2010ApJ...718..184C}). We primarily used the \cite{2020ARA&A..58..529S} parameterization for our analysis to facilitate comparisons with similar studies. Additionally, we employed the more flexible \cite{2008ApJ...685.1046L} parameterization in the spectral energy distribution (SED) fitting procedure to recover the true shape of the attenuation curve.

Studies have investigated trends that connect the shape of the attenuation and/or extinction curves to fundamental galaxy properties. The diverse curve shapes are often attributed to primary underlying factors, including RT effects (e.g., \citealp{1999ApJS..123..437F, 2000ApJ...528..799W, 2016ApJ...833..201S}) and intrinsic dust properties (e.g., \citealp{2018MNRAS.478.2851M, 2020MNRAS.492.3779H, 2022MNRAS.517.2076M}). Although valuable information was offered about the primary drivers of the diversity of the attenuation curves (for review, see \citealp{2020ARA&A..58..529S}), the study results and interpretation often contradict each other, probably because the galaxy samples and methods that were used to characterize dust attenuation differ. 

A common finding among the different results is an anticorrelation between the UV-optical slope $S$ and the optical attenuation $A_V$, where galaxies with high (low) $A_V$ typically exhibit flatter (steeper) slopes. This trend was observed at low ($z \sim 0$; \citealp{1994ApJ...429..582C, 2000ApJS..129..147C, 2017ApJ...837..170L, 2018ApJ...859...11S, 2019MNRAS.486..743D, 2020ARA&A..58..529S, 2023ApJ...957...75Z, 2025arXiv250415346M}), intermediate ($z \sim 0.1-3.5$; \citealp{2013A&A...558A..67A, 2016ApJ...827...20S, 2020ApJ...888..108B}), and high redshift ($z > 4$; \citealp{2022A&A...663A..50B, 2024NatAs.tmp...20M, 2025MNRAS.539..109F}), and it aligns with RT predictions that suggest that shallower curves emerge with increasing dust column density for which $A_V$ serves as a proxy (\citealp{2005MNRAS.359..171I, 2013MNRAS.432.2061C, 2016ApJ...833..201S, 2020MNRAS.491.3937T, 2021MNRAS.507.2755L}). 

After $A_V$ is accounted for, most trends between the slope and inclination, stellar mass ($M_*$), the specific star formation rate (sSFR), and metallicity ($Z$) become mostly insignificant (\citealp{2020ARA&A..58..529S}).
For example, a negative correlation of the slope with inclination was observed (\citealp{2011MNRAS.417.1760W, 2017ApJ...851...90B}) and predicted (\citealp{2005MNRAS.359..171I, 2013MNRAS.432.2061C, 2020MNRAS.491.3937T, 2025arXiv250213240S}). It is probably caused by the increased dust opacity (i.e., $A_V$) in edge-on systems. 
On the other hand, the correlation of the slope with $M_*$ varies widely in the studies. Some found negative trends (\citealp{2018ApJ...859...11S, 2018ApJ...853...56R, 2019A&A...630A.153A, 2020ARA&A..58..529S, 2020ApJ...899..117S, 2025arXiv250415346M}), others reported positive trends (\citealp{2012A&A...545A.141B, 2015ApJ...814..162Z}), and some detected no significant trends at all (\citealp{2007ApJS..173..392J, 2016ApJ...827...20S, 2016ApJ...818...13B, 2017ApJ...840..109B, 2022ApJ...932...54N, 2025MNRAS.539..109F}). 
Likewise, studies of the correlation of the slope with sSFR and/or stellar age (an inverse of the sSFR can be considered as a proxy for the mean stellar age; e.g., \citealp{2015ApJ...814..162Z, 2024arXiv241014671L}). reported mixed results. Some reported negative $S$-sSFR or positive $S$-age trends (\citealp{2013ApJ...775L..16K, 2015ApJ...814..162Z, 2019MNRAS.488.2301T, 2020ApJ...888..108B}), while others reported inverse trends (\citealp{2007A&A...472..455N, 2009A&A...499...69N, 2018ApJ...859...11S}) or no significant trends (\citealp{2007ApJS..173..392J, 2011MNRAS.417.1760W, 2016ApJ...818...13B, 2017ApJ...840..109B, 2019ApJ...872...23S}).
Similarly, the correlation of the slope with metallicity was reported to vary, with either negative trends (\citealp{2018ApJ...853...56R, 2020ApJ...899..117S, 2020ApJ...903L..28S}) or a lack of trends (\citealp{2017ApJ...840..109B, 2018ApJ...859...11S, 2025MNRAS.539..109F}). 

Some works identified the characteristic UV bump amplitude at rest $\sim 2175\AA$ ($B$), and explored the correlations between $B$ and the fundamental galaxy properties, which yielded varied results. In general, $B$ appears to decrease with increasing $A_V$ in attenuation curves of low- and intermediate-$z$ galaxies (e.g., \citealp{2018ApJ...859...11S, 2019MNRAS.486..743D, 2019MNRAS.488.2301T}). This agrees with RT models, which show that the UV bump is suppressed in dustier systems (\citealp{2004ApJ...617.1022P, 2016ApJ...833..201S, 2025A&A...695A..77D}). Other studies reported no significant $B-A_V$ trends, however (\citealp{2020ApJ...888..108B, 2023ApJ...957...75Z}).
Additionally, studies of the UV bump strengths in the extinction curves of local galaxies (\citealp{2000ApJS..129..147C, 2003ApJ...594..279G, 2024ApJ...970...51G}),  distant quasars (\citealp{2006MNRAS.367..945Y, 2015MNRAS.454.1751M}), and GRB afterglows (\citealp{2011A&A...532A.143Z, 2018MNRAS.479.1542Z}) reported opposite $B$-$A_V$ trends that probably reflect variations in the dust properties. 

The trend of $B$  with other galaxy properties is also mixed. For example, UV bumps are more prominent in edge-on systems (\citealp{2011MNRAS.417.1760W, 2013ApJ...775L..16K, 2017ApJ...851...90B}), although some studies reported no significant link to inclination (\citealp{2020ApJ...888..108B, 2023ApJ...957...75Z}). 
Next, $B-M_*$ trends are only marginally significant (\citealp{2007A&A...472..455N, 2020ApJ...888..108B, 2025MNRAS.539..109F}). They have a (weakly) positive (\citealp{2012A&A...545A.141B, 2021ApJ...909..213K}) or negative correlation in various cases (\citealp{2020ApJ...903..146B}). 
Moreover, $B$ generally decreases with increasing sSFR (\citealp{2011MNRAS.417.1760W, 2012A&A...545A.141B, 2013ApJ...775L..16K, 2021ApJ...909..213K, 2023ApJ...957...75Z, 2025PASA...42...22B}), or decreasing stellar age (\citealp{2012A&A...545A.141B}). Some studies found no significant trends (\citealp{2020ApJ...888..108B}) or inverse trends with sSFR (\citealp{2009A&A...499...69N, ormerod2025detection2175aauvbump}) and age (\citealp{2019MNRAS.488.2301T}), however.
UV bumps also tend to be more prominent in high-metallicity systems (\citealp{2020ApJ...899..117S, 2022MNRAS.514.1886S}), although this trend is still debated (\citealp{2018ApJ...859...11S, 2025PASA...42...22B, 2025MNRAS.539..109F}). 

Finally, a positive correlation between $S$ and $B$ was predicted by the RT models (\citealp{2016ApJ...833..201S, 2018ApJ...869...70N}) and confirmed by observations at $z \sim 0$ (\citealp{2018ApJ...859...11S, 2020ARA&A..58..529S}) and at intermediate $z$ (\citealp{2013ApJ...775L..16K, 2018MNRAS.475.2363T, 2019MNRAS.488.2301T}). Other studies reported no significant trends (\citealp{2011A&A...533A..93B, 2012A&A...545A.141B}) or even inverse $S-B$ trends in attenuation curves (\citealp{2020ApJ...899..117S}) and extinction curves (\citealp{1989ApJ...345..245C, 2024ApJ...970...51G}), however. This might indicate that the intrinsic dust properties are the main factor that shapes these curves. 

In summary, except for the $S-A_V$ relation, there is no broad consensus on independent correlations between the parameters of the dust attenuation law and other galaxy properties. The unclear trends might reflect the complex interplay of the dust size distribution and composition and RT effects.

\cite{2024NatAs.tmp...20M} analyzed the dust attenuation curves of 173 galaxies at $z \sim 2-12$, inferred from our customized SED fitting model and spectroscopic observations with the {\it James Webb} Space Telescope (JWST). The focus of that work was the redshift evolution on the UV-optical slope (S) and UV bump strength (B)\footnote{A subset of the galaxy properties specifically connected to the results presented by \cite[][e.g., redshift, $S$, $B$, $c_4$, and $A_V$]{2024NatAs.tmp...20M} was made publicly available upon the publication of \cite{2024NatAs.tmp...20M}.}, expanding on similar studies at intermediate (e.g., \citealp{2011A&A...533A..93B, 2012A&A...545A.141B, 2022MNRAS.513.4431B}) and high $z$ (\citealp{2010A&A...523A..85G, 2015ApJ...800..108S}). In this work, we revisit this sample (Sect. \ref{Data}) and adopt the same method as \cite{2024NatAs.tmp...20M} (Sect. \ref{Method}) to systematically explore the trends between $S$, $B$, and remaining galaxy properties inferred from the SED fitting or emission line diagnostics (Sect. \ref{Results}). We determine the main trends between the fundamental galaxy properties in Sect. \ref{global_trends}. 
Finally, we discuss the implications for the physical processes that might drive the correlations between (i) $S$ and stellar age (${\langle a \rangle}_*^{\rm{m}}$) and (ii) between $B$ and the oxygen abundance ($12 + \log(\rm{O/H})$) in Sect. \ref{Discussion}.

\section{Spectroscopic data} \label{Data}

This section provides a brief overview of the JWST/NIRSpec data, the sample selection criteria, the galaxy dataset, and the slit-loss correction method. Full details can be found in the original paper of \cite{2024NatAs.tmp...20M}. 

We analyzed the JWST Near Infrared Spectrograph (NIRSpec) prism/clear (henceforth prism) spectroscopic dataset from the DAWN JWST Archive (DJA)\footnote{\url{https://dawn-cph.github.io/dja/index.html}}. This archive includes uniformly reduced NIRSpec prism data from various JWST surveys, reprocessed by the Cosmic Dawn Center, with the MSAEXP Python script (\citealp{Brammer_msaexp_NIRSpec_analyis_2022, 2022zndo...7299500B, 2024Sci...384..890H, 2025A&A...697A.189D}). The low-resolution ($R \sim 30-300$) 1D prism spectra span a broad wavelength range ($\lambda \sim 0.6-5.3 \mu m$), but the limited velocity resolution ($\gtrsim 1000\ \mathrm{km \ s^{-1}}$) leaves nebular emission lines unresolved (\citealp{2022A&A...661A..80J}). We masked wavelengths below the rest-frame Ly$\alpha$ to avoid contamination from the neutral intergalactic medium (IGM). 

To account for flux losses due to slit effects, we first extracted photometric data from the \texttt{grizli-v7.0} photometric catalogs using the publicly available Python script provided by the DJA\footnote{\url{https://dawn-cph.github.io/dja/blog/2023/07/14/photometric-catalog-demo/}}. The photometry spans multiple bands from the {\it Hubble Space Telescope} Wide Field Camera 3 (WFC3) and Advanced Camera for Surveys (ACS), and JWST Near Infrared Camera (NIRCam) and Near Infrared Imager and Slitless Spectrograph (NIRISS) instruments. The exact number of bands we used varied by field and typically ranged from 16 to 30 (see Table \ref{bands}). We used the flux measurements from aperture 1 (corresponding to a diameter of 0.5 arcseconds), which were corrected for flux losses outside the aperture.

We overlaid the photometric points on the spectra and then corrected the spectrum by dividing it by one of the throughputs provided in the JWST user documentation\footnote{\url{https://jwst-docs.stsci.edu/jwst-near-infrared-spectrograph/nirspec-operations/
nirspec-mos-operations/nirspec-mos-operations-slit-losses}} until we achieved a satisfactory alignment between the spectra and photometry. Furthermore, to assess potential systematics in the above visually based method, we applied an independent correction for slit losses. This approach estimates the offsets between spectroscopic and photometric flux densities, fits these offsets with an uncertainty-weighted linear relation, and then corrects the spectra accordingly.
Finally, we compared the SED fitting results on spectra with and without slit-loss corrections to ensure consistency. Our key findings on the redshift evolution of dust attenuation curves remain robust across all methods (see \citealp{2024NatAs.tmp...20M} for more details).

For our sample selection, we first identified spectra with reliable spectroscopic redshifts, based on robust spectroscopic features (grade 3 spectra). We then focused on galaxies at $z > 2$ to probe the rest-frame UV-optical range, enabling accurate constraints on the shape of the dust attenuation curve. We refined our sample by selecting spectra with a high signal-to-noise ratio S/N $ > 3$) per channel in the rest-frame $1925-2425~\AA$ wavelength range, corresponding to the UV bump region. We finally restricted the sample to galaxies with significant dust attenuation along the LOS, that is, to galaxies with $A_V > 0.1$ at  $>3\sigma$ significance. This yielded a final sample of 173 dusty galaxies (see supplementary material of \citealp{2024NatAs.tmp...20M}) that covered a broad redshift range ($2.0 \lesssim z \lesssim 11.4$) and spanned a wide range of fundamental galaxy properties in terms of stellar masses ($6.9 \lesssim \log{M_*/M_{\odot}} \lesssim 10.9$), star formation rates ($0.1 \lesssim \rm{SFR}/M_{\odot} \ \rm{yr^{-1}} \lesssim 240$), and $V$-band dust attenuation ($0.1 \lesssim A_V \lesssim 2.3$).

\begin{figure*}
\centering
\includegraphics[width=0.56\hsize]{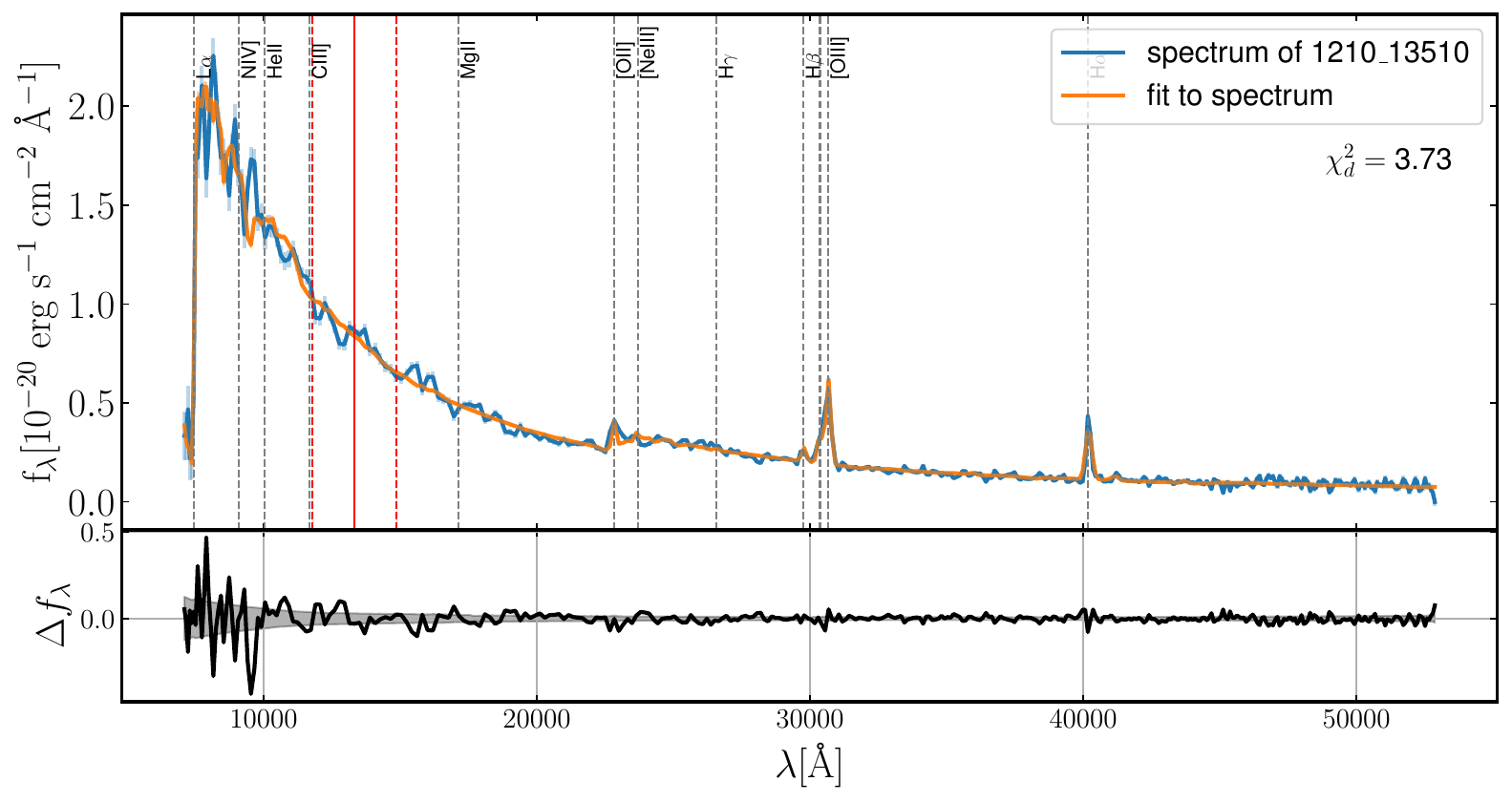}
\includegraphics[width=0.38\hsize]{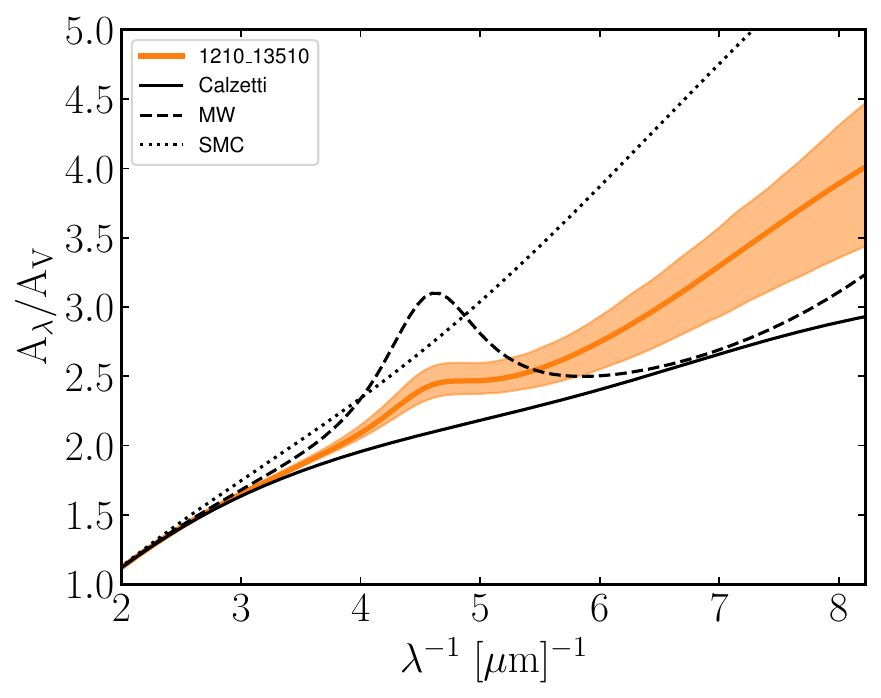}
 \caption{Example of the SED fitting results for a galaxy with a UV-bump detection ($1210\_13510$ at $z \approx 5.12$). Left panel: NIRSpec prism spectrum (blue) and best-fit posterior model (orange; top panel). The shaded regions represent the respective $1\sigma$ uncertainties. The vertical gray and red lines indicate the positions of potential emission lines and the UV bump absorption feature, respectively. The residuals of the best-fit model relative to the observed spectrum ($\Delta f_{\lambda}$) are reported in the bottom panel, along with the 1$\sigma$ uncertainties. Right panel: Best-fit dust attenuation curve with $1\sigma$ uncertainties derived from a bootstrap method that generated 5000 attenuation curves by randomly sampling the $c_1-c_4$ parameters (Eq. \ref{dust_law}) from the posterior distribution. For comparison, the Calzetti, MW, and SMC empirical curves are shown as solid, dashed, and dotted black lines, respectively.
 }
 \label{spec}
\end{figure*}

\section{Method} \label{Method}

In this section, we provide a brief overview of our customized SED fitting method in Sect. \ref{method_SED}, which is previously detailed by \cite{2023A&A...679A..12M}. In Sect. \ref{method_metallicity} we present our new analysis of oxygen abundance measurements using the direct $T_e$ (electron temperature) method, which is fully described by \cite{2024ApJ...962...24S}.

\subsection{SED fitting with BAGPIPES} \label{method_SED}

We employed a customized version \citep{2023A&A...679A..12M} of the \texttt{BAGPIPES} SED fitting code (\citealp{2018MNRAS.480.4379C}), which uses a Bayesian framework and the \texttt{MultiNest} nested-sampling algorithm (\citealp{2019OJAp....2E..10F}) to construct realistic galaxy spectra. The stellar emission was modeled using the stellar population synthesis (SPS) models of \cite{10.1093/mnras/stw1756}. In \texttt{BAGPIPES}, nebular continuum and line emissions are modeled using precomputed grids of \texttt{CLOUDY} models (\citealp{2017RMxAA..53..385F}). In these grids, the ionization parameter varies in the range of $\log{U} \sim -4, 0$ by changing the number of hydrogen-ionizing photons $Q_H$ and assuming a constant hydrogen density of $n=$100 atoms $\rm{cm^{-3}}$. The grids are interpolated based on the stellar population age ($a_* \sim 1 \ \rm{Myr} - a_{\rm{cosmo}}$, where $ a_{\rm{cosmo}}$ is the age of the Universe) and metallicity ($Z \sim 0.005 - 5. \ Z_{\odot}$) to derive the emission line strengths and nebular continuum of the output spectrum. 

Our fiducial star formation history (SFH) model was the nonparametric SFH model with a continuity prior (\citealp{2019ApJ...876....3L}) because it favors flexibility in capturing more accurate SFH and unbiased physical property estimates than parametric models. \texttt{BAGPIPES} accounts for the transmission of galaxy emission through the neutral gas in the IGM, following \cite{2014MNRAS.442.1805I}. \texttt{BAGPIPES} also includes warm dust emission from \ion{H}{II} regions computed with \texttt{CLOUDY}, along with cold dust emission in the interstellar medium (ISM), modeled as a graybody. The dust attenuation incorporates a two-component dust screen model to account for additional attenuation experienced by young stars within their stellar birth clouds (\citealp{2000ApJ...539..718C}). 

For the fiducial dust attenuation model, we integrated the analytical model from \cite{2008ApJ...685.1046L} into \texttt{BAGPIPES}. The flexible attenuation model effectively recovers local empirical templates (Calzetti, MW, and SMC, as demonstrated in \citealp{2023A&A...679A..12M}) and can capture variations in the dust attenuation laws in galaxies that are expected at higher redshifts (\citealp{2023A&A...679A..12M, 2024NatAs.tmp...20M}, see also \citealp{2023Natur.621..267W, 2024arXiv240805273S}). This is particularly relevant for high-redshift galaxies because there is no physical basis for assuming that their dust attenuation curves mirror those of local galaxies. The analytical form of the normalized dust attenuation is defined as

\begin{equation} \label{dust_law}
\begin{split}   
A_{\lambda}/A_V &= \frac{c_1}{(\lambda/0.08)^{c_2}+(0.08/\lambda)^{c_2} +c_3} \\ 
& +  \frac{233[1-c1/(6.88^{c_2}+0.145^{c_2}+c_3)-c_4/4.60]}{(\lambda/0.046)^2+(0.046/\lambda)^2+90} \\ 
& + \frac{c_4}{(\lambda/0.2175)^2+(0.2175/\lambda)^2-1.95},
\end{split}
\end{equation}%
where $c_1 - c_4$ are dimensionless parameters, and $\lambda$ is the wavelength in $\mu m$. 

To assess the difference in model performance between our fiducial SED model with the flexible dust attenuation approach and the SED models with the Calzetti curve, we used $\chi^2$ statistics. In $\sim 86\%$ of galaxies, the fiducial model yielded a statistically significant improvement (with $P<0.05$), while in $\sim 13\%$ of cases, the improvement was present, but not statistically significant ($0.05<P<1$). Only $1\%$ of the sources showed no significant difference between the models, probably because their attenuation curves are intrinsically flat. Additionally, we conducted a similar analysis by comparing the fiducial model to a bump-free version of the same model (with $c_4 = 0$) for a subsample of 28 galaxies with a detectable UV bump. In this case, we found that the fiducial model provided a significantly better fit for $ \sim 86\%$ of the sources ($P<0.05$; see \citealp{2024NatAs.tmp...20M}).
Fig. \ref{spec} depicts an example of SED fitting results for a high-$z$ source with a detected UV bump with an amplitude of $B = 0.11 \pm 0.04$ ($\sim 0.31 B^{\rm{MW}}$) or $c_4 = 0.014 \pm 0.004$.

\begin{figure*}
\centering
\includegraphics[width=\hsize]{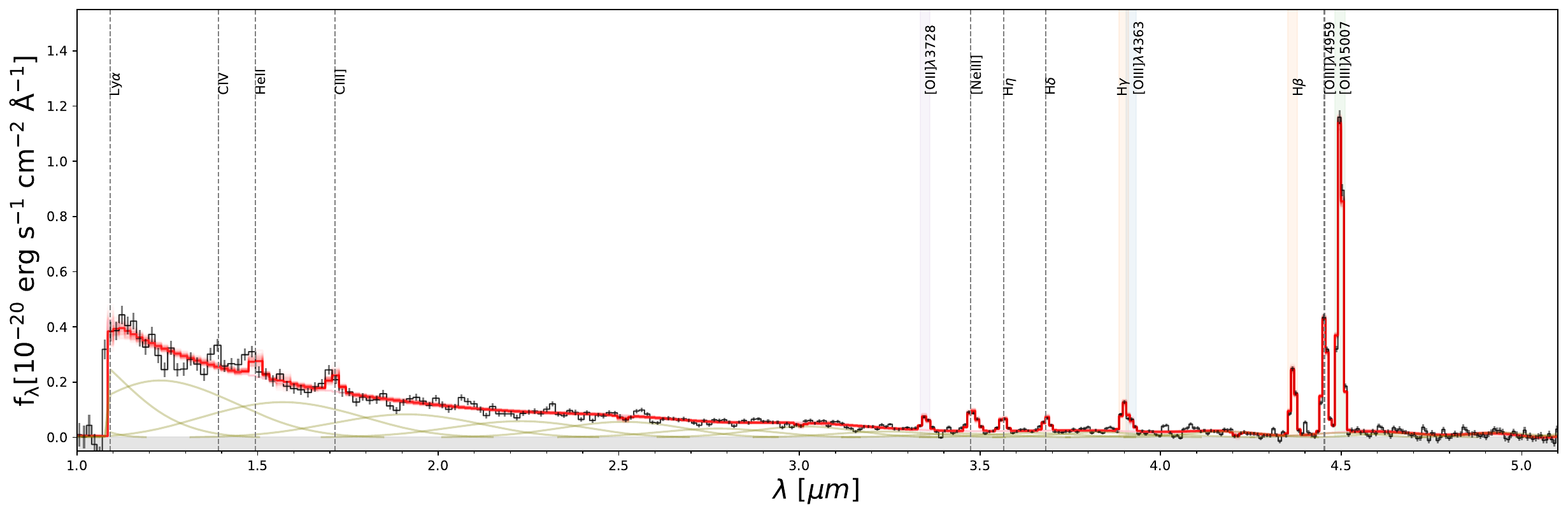}
 \caption{Example of SED fitting results for a galaxy with oxygen abundance measurements ($1210\_5173$ at $z \approx 7.99$). The NIRSpec prism spectrum (black) and the best-fit model (red) are depicted with their respective $1\sigma$ uncertainties. 
 The vertical lines indicate the positions of key optical emission lines we used to estimate the oxygen abundance: $\rm{H\gamma}$ and $\rm{H\beta}$ (orange), $[\ion{O}{II}] \lambda3728$ (purple), $[\ion{O}{III}] \lambda4363$ (blue), and $[\ion{O}{III}] \lambda5007$ (green). The vertical dashed gray lines indicate the positions of the remaining emission lines. The olive lines indicate the continuum emission levels.
 }
 \label{spec_OH}
\end{figure*}

\subsection{Oxygen abundance measurements} \label{method_metallicity}

We derived the oxygen abundances using the direct $T_e$ (electron temperature) empirical metallicity calibrations for high-redshift ($z \sim 2-9$) low-metallicity ($12 + \log(\rm{O/H}) \sim 7.0-8.4$) galaxies, following \cite{2024ApJ...962...24S}, and employing the software package \texttt{PyNeb} (\citealp{2015A&A...573A..42L}). 

We fit the galaxy spectra using the DJA NIRSpec data products code (\citealp{2024Sci...384..890H, 2025A&A...697A.189D}) to derive the emission line fluxes ($F$) and the corresponding uncertainties ($F_{\rm{err}}$) for the full galaxy sample. To ensure reliable oxygen abundance estimates, we restricted our analysis to galaxies with bright ($F/F_{\rm{err}} > 3$) optical emission lines that we used as metallicity calibrators, namely $\rm{H\alpha}$ (or $\rm{H\gamma}$ for $z \gtrsim 7$), $\rm{H\beta}$, $[\ion{O}{III}] \lambda5007$, $[\ion{O}{III}] \lambda4363$, and $[\ion{O}{II}] \lambda3728$ (due to the NIRSpec prism resolution, we fit the blended
$[\ion{O}{II}] \lambda \lambda3727, 3729$ emission lines as a single $[\ion{O}{II}] \lambda3728$ line). This yielded a subset of 14 sources at $ z\sim4.6-8.0$. 

We corrected the line fluxes for nebular reddening, $E(B-V)_{\rm{gas}}$, estimated from the observed-to-intrinsic ratios of the hydrogen Balmer recombination lines. This was typically $\rm{H\alpha/H\beta}$ for $z \lesssim 7$ galaxies and $\rm{H\gamma/H\beta}$ for $z \gtrsim 7$ sources. The intrinsic Balmer line ratios and the correction for reddening estimates assumed $T_e = 15000 \ \rm{K}$, an electron density of $n_e = 300 \ \rm{cm^{-3}}$ (\citealp{Strom_2017, 2024ApJ...962...24S}), and a Calzetti-like attenuation curve (\citealp{2000ApJ...533..682C}), which is overall consistent with the empirically measured attenuation curve of high-$z$ ($z\sim 4.5-11.5$) sources (\citealp{2024NatAs.tmp...20M}). The reddening correction does not account for possible stellar absorption below the Balmer lines. While we did not apply explicit corrections for stellar absorption, our sample consists of bursty star-forming high-redshift galaxies with strong emission lines, where the effect of stellar absorption is expected to be weak. For instance, \cite{2015ApJ...806..259R} reported typical absorption corrections of $\lesssim 10\%$ for Balmer emission line fluxes in high-$z$ galaxies. When these corrections are applied to our data, we obtain systematically lower $E(B-V)_{\rm{gas}}$ values that are generally within $1\sigma$ of our original estimates.

Next, we estimated the electron temperature for the high-ionization region, $T_e([\rm{O^{2+}}])$, using the $[\ion{O}{III}] \lambda4363/[\ion{O}{III}] \lambda5007$ ratio and assuming an electron density of $n_e = 300 \ \rm{cm^{-3}}$ (\citealp{Strom_2017, 2024ApJ...962...24S}) because the density-sensitive emission lines ($[\ion{O}{II}] \lambda \lambda3727, 3729$ and $[\ion{S}{II}] \lambda \lambda6718, 6733$) are blended. For the low-ionization zone, $T_e([\rm{O^{+}}])$, we adopted the relation $T_e([\rm{O^{+}}]) = 0.7 \times T_e([\rm{O^{2+}}]) + 3000 \ \rm{K}$ (\citealp{10.1093/mnras/223.4.811}).

Ionic abundances ($\rm{O^{2+}/H^+}$ and $\rm{O^{+}/H^+}$) were calculated from the $[\ion{O}{III}] \lambda5007/\rm{H\beta}$ and $[\ion{O}{II}] \lambda3728/\rm{H\beta}$ ratios using inferred electron temperatures $T_e([\rm{O^{2+}}])$ and $T_e([\rm{O^{+}}])$, respectively. 
The total oxygen abundance was then derived as $\rm{O/H} = \rm{O^{2+}/H^+} + \rm{O^{+}/H^+}$. 

The uncertainties were estimated by randomly drawing flux values from a Gaussian distribution with the same $1\sigma$ as the line flux uncertainties for each source and by recalculating $E(B-V)_{\rm{gas}}$, $T_e$, and $\rm{O/H}$ for each realization. This procedure was performed 500 times, and the $1\sigma$ uncertainties were derived from the $16^{th}- 84^{th}$ percentile range of the resulting distribution.  

The oxygen abundances of our galaxy subsample lie in the range of $12 + \log(\rm{O/H}) \sim 7.7-8.7$, and one outlier reaches $12 + \log(\rm{O/H}) \sim 9.5$. Because the direct $T_e$ method is only valid for low-metallicity sources, we excluded this outlier from our final subset. 
This yielded our final subset of 13 sources with reliable oxygen abundances. It covered a redshift range of $4.6 \lesssim z \lesssim 8.0$, stellar masses of $7.8 \lesssim \log{M_*/M_{\odot}} \lesssim 9.1$, and a $V$-band dust attenuation $0.2 \lesssim A_V \lesssim 1.4$. 
An example fit to the spectrum of a high-$z$ galaxy (ID 1210\_5173) with an oxygen abundance measurement ($12 + \log(\rm{O/H}) \sim 8.09 \pm 0.09$) is shown in Fig. \ref{spec_OH}. 

In Fig. \ref{spec_OH}, the $\rm{H\gamma}$ and $[\ion{O}{III}] \lambda4363$ lines appear to be partially blended in the low-resolution NIRSpec prism spectrum because of their close separation ($\sim 22 \AA$), as shown in Fig. \ref{spec_OH}. The two lines were fit independently, where each line was modeled as a single-Gaussian profile centered at the line wavelength, and convolved with the instrument line spread function (\citealp{2024Sci...384..890H, 2025A&A...697A.189D}). This degeneracy may introduce additional uncertainties in the individual flux measurements of the two lines. We verified, however, that excluding these sources (two $z > 7$ sources for which $E(B-V)_{\rm{gas}}$ was calculated from the $\rm{H\gamma/H\beta}$ ratio) from our subsample does not alter the observed $B-12 + \log(\rm{O/H})$ trend, which remains significant.

For comparison, we also estimated oxygen abundances using the strong-line diagnostics of \cite{Nakajima_2022}. This method combines two optical emission line ratios,
\begin{equation}
  \rm{R23} = \frac{[\ion{O}{III}] \lambda4959, 5007 +[\ion{O}{II}] \lambda3727, 3730}{\ion{H}{$\beta$}},  
\end{equation}%
and 
\begin{equation}
    \rm{O32} = \frac{[\ion{O}{III}] \lambda5007}{[\ion{O}{II}] \lambda3727, 3730},
\end{equation}
to provide an independent metallicity measurement.
For low-metallicity sources ($\rm{O32} \geq 2$), we used $\rm{R23}$, incorporating calibrations on the equivalent width of the \ion{H}{$\beta$} line, while for the high-metallicity sources ($\rm{O32} < 2$), we instead used $\rm{O32}$ (\citealp{Nakajima_2022, 2025ApJ...988...26W}). 

The oxygen abundances derived from the two methods are broadly consistent, but the direct $T_e$ method systematically yields higher oxygen abundances (up to a factor of $\sim 1.1$) than the calibration by \cite{Nakajima_2022}.

\begin{figure}
\centering
\includegraphics[width=\hsize]{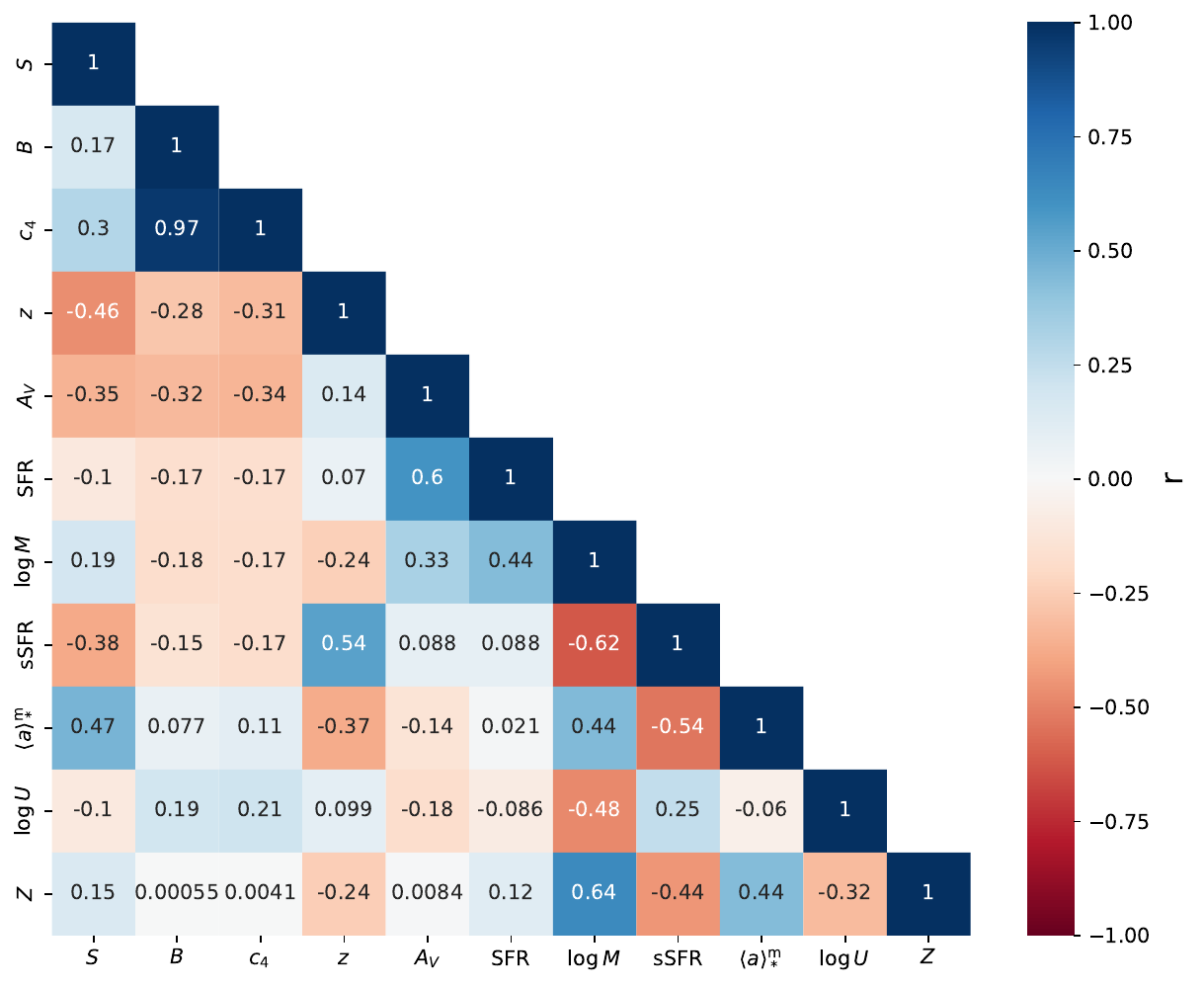}
\caption{Correlation matrix heatmap for the dust attenuation curve properties (the UV-optical slope $S$ and the UV bump amplitude, i.e., $B$ and $c_4$) and fundamental galaxy parameters as inferred from the SED fitting: Redshift ($z$), $V$-band attenuation ($A_V$), SFR, stellar mass ($\log{M}$), sSFR, stellar age (${\langle a \rangle}_*^{\rm{m}}$), ionization parameter ($\log{U}$), and metallicity ($Z$). The color-code represents Pearson correlation coefficient ($r$) values, which quantify the strength and the sign of the correlation. Typically, $|r| = 1$ signifies a perfect correlation,  $|r| = 0.7-1$ a very strong correlation,  $|r| = 0.5-0.7$ a strong correlation, and $|r| \sim 0.3-0.5$ a moderate correlation (\citealp{Benesty2009}). 
\label{PCA}
}
\end{figure}

\section{Trends with the slope (S) and UV bump (B)}\label{Results}

In \cite{2024NatAs.tmp...20M} we showed that attenuation curves tend to flatten at high redshift, and we discussed the origin of this trend. In this follow-up work, we explore potential trends between the shape of the attenuation curves and the main galaxy properties as inferred from the SED fitting: $V$-band attenuation ($A_V$), star formation rate (SFR) averaged over 10 Myr, stellar mass (${M_*}$), sSFR ($\rm{sSFR = SFR}/M_{*}$), mass-weighted stellar age (${\langle a \rangle}_*^{\rm{m}}$), ionization parameter ($\log{U}$), and metallicity ($Z$). For the metallicity, we also considered the more reliable estimates of the oxygen abundance ($12 + \log(\rm{O/H})$) obtained from optical emission lines. We used the parameterization by \cite{2020ARA&A..58..529S} ($S$ and $B$ parameters) for the dust attenuation curve.  The final goal was to separate the physical drivers of the potential trends. 

To identify independent trends between dust attenuation curve properties and global galaxy parameters, the ideal approach would involve slicing the sample across multiple dimensions simultaneously, that is, analyzing one variable while holding all other properties fixed. Our relatively small sample of 173 high-redshift galaxies limits such an analysis because of the low number statistics in multiparameter bins, however. Therefore, our conclusions are inherently subject to sample size limitations, and future studies based on larger and more homogeneous datasets are essential to verify and refine the trends we identified here. To mitigate these limitations, we augmented our analysis by adopting two independent statistical methods, namely the Pearson correlation analysis and the random forest regression (RFR).

The correlation matrix offers a convenient first step for identifying linear relations between individual parameter pairs. We computed the Pearson correlation coefficient ($r$) to quantify the linear correlation between $S$, $B$, and all the fundamental properties inferred from the SED fitting. For comparison with previous studies, we also included the sSFR. We claim that two properties are correlated when $|r|>0.3$ (\citealp{Greenacre}; see the caption of Fig. \ref{PCA} for further details).

The correlation matrix is reported in Fig. \ref{PCA} and shows the following main results:
\begin{itemize}
\item We confirm the $S-z$ ($r=-0.46$) and $B-z$ ($r=-0.28$), or equivalently, $c_4-z$ ($r=-0.31$), anticorrelations found by \cite{2024NatAs.tmp...20M}. 
\item We find that $S$ is anticorrelated with the V-band attenuation, $A_V$ ($r = -0.35$) and sSFR ($r = -0.38$) and correlated with stellar age, ${\langle a \rangle}_*^{\rm{m}}$ ($r = 0.47$). 
\item The UV bump amplitude $B$ (or $c_4$) is anticorrelated with $A_V$ ($r = -0.32$ for $B$, $r = -0.34$ for $c_4$).
 \end{itemize}

 The main limitation of the correlation matrix analysis is that it estimates linear relations between pairs of parameters without accounting for the influence of other variables (\citealp{Benesty2009}). To complement and refine this analysis, we therefore considered the RFR (\citealp{Benesty2009}) method and exploited the \texttt{scikit-learn} library (\citealp{scikit-learn}).
The RFR is an ensemble-learning method that constructs multiple decision trees and combines their results to improve the prediction accuracy. Unlike the correlation matrix, which evaluates each pair of parameters in isolation, RFR can capture nonlinear relations between the input parameters and the combined influence of all input parameters on a target variable. It provides a measure of feature importance, indicating the contribution of each input parameter to reducing the overall prediction error on the target variable.
We used the RFR to assess the importance of various global galaxy properties (e.g., $z$, $A_V$,  ${\langle a \rangle}_*^{\rm{m}}$, and sSFR) in predicting $S$ and $B$. 

\begin{figure}
\centering
\includegraphics[width=\hsize]{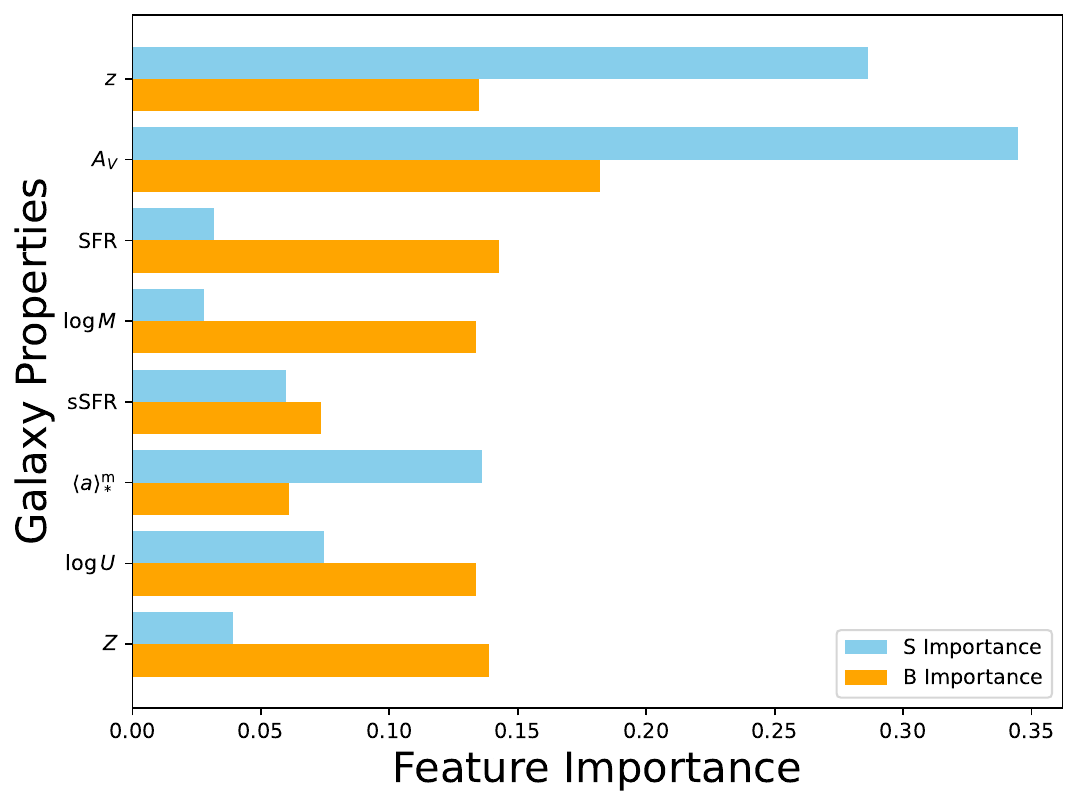}
\caption{Feature importance resulting from the RFR method. The score assigned to each galaxy property quantifies its importance in predicting the slope ($S$, horizontal sky blue bars) and the UV bump strength ($B$, orange bars).
\label{RFR}
}
\end{figure}

Fig. \ref{RFR} shows the feature importance of each galaxy property in predicting $S$ and $B$. The most important predictors of $S$ are $z$ ($\sim 0.29$), $A_V$ ($\sim 0.34$) and ${\langle a \rangle}_*^{\rm{m}}$ $\sim 0.14$, , while sSFR has a lower feature importance ($\sim 0.06$). This might be explained by the strong correlations of sSFR with $z$ ($r = 0.54$) and ${\langle a \rangle}_*^{\rm{m}}$ ($r = -0.54$), making the S-sSFR correlation redundant when $z$ and ${\langle a \rangle}_*^{\rm{m}}$ are included in the model.
The interpretation of the results for $B$ is more complicated: The RFR confirms that $A_V$ is the most important quantity that shapes the UV bump strength (feature importance ($\sim0.2$)), but properties such as $z$, SFR, $M_*$ and $U$ still provide an important contribution ($\sim 0.1-0.15$).

\begin{figure*}
\centering
\includegraphics[width=0.49\hsize]{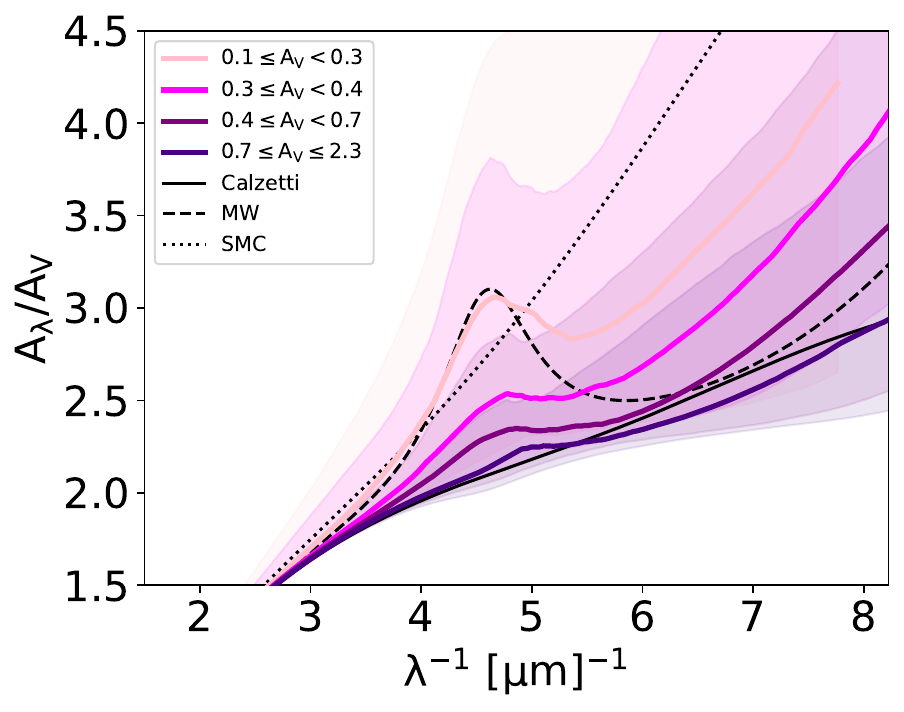}
\includegraphics[width=0.49\hsize]{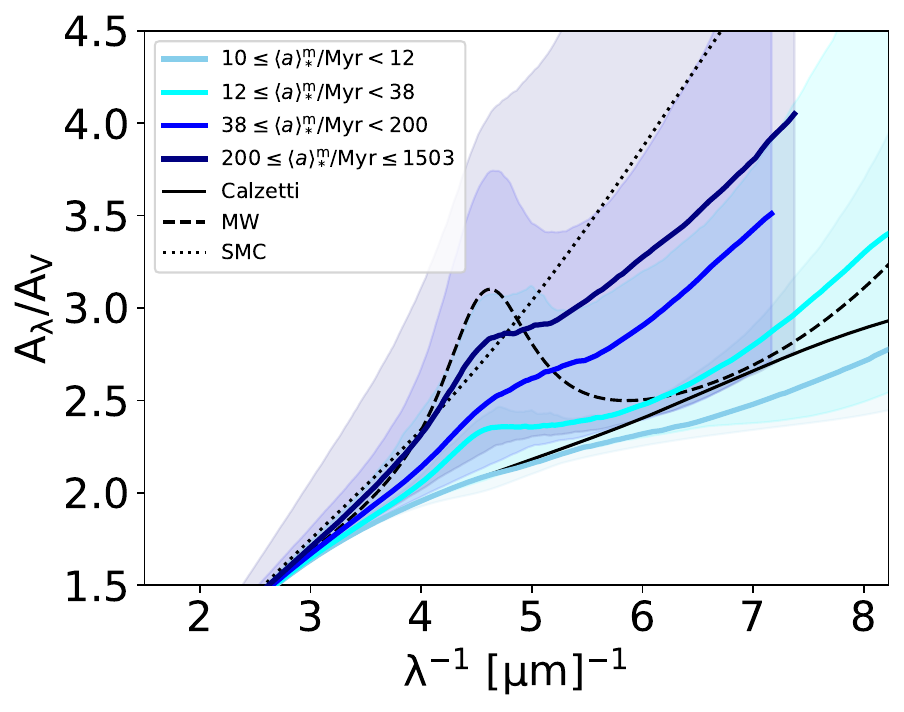}
\caption{Median dust attenuation curves for our full galaxy sample binned in terms of key parameters: (left panel) $V$-band attenuation ($A_V$), and (right panel) mass-weighted stellar age (${\langle a \rangle}_*^{\rm{m}}$).  
The median attenuation curves are plotted as solid lines, and the shaded regions indicate their $1\sigma$ dispersion. These curves and uncertainties were estimated using a bootstrapping approach, where 5000 synthetic curves were generated and the $16^{th}$, $50^{th}$, and $84^{th}$ percentiles were taken from the resulting distribution. For comparison, the Calzetti, MW, and SMC empirical curves are displayed as solid, dashed, and dotted black lines, respectively. The bin limits were chosen to maintain a roughly equal number of sources per bin ($\sim 40-50$ sources in four bins).
\label{trends}
}
\end{figure*}

Fig. \ref{trends} shows the median dust attenuation curves along with the corresponding $1\sigma$ dispersion for galaxy subsamples binned in terms of the two key properties shaping the dust attenuation curve: $A_V$ and ${\langle a \rangle}_*^{\rm{m}}$ (left and right panels, respectively). We split our sample into four bins per property, ensuring an approximately equal number of sources in each bin ($\sim 40-50$ sources). A first qualitative inspection shows that the slope ($S$) of the dust attenuation curves is anticorrelated with $A_V$, but is correlated with ${\langle a \rangle}_*^{\rm{m}}$. 

In the next sections, we briefly discuss the $S-z$ and $B-z$ trends detailed by \cite{2024NatAs.tmp...20M} (Sect. \ref{curve_z_trend}), and we focus on the $S-A_V$ and $B-A_V$ trends (Sect. \ref{curve_Av_trend}) and $S-{\langle a \rangle}_*^{\rm{m}}$ trends (Sect. $\ref{S_age_trend}$). 
 For the galaxy subset with determined oxygen abundances from the optical emission lines, we further searched for a potential $B-\rm{12+ \log{O/H}}$ correlation (Sec. \ref{S_Z_trend}).

A key caveat is that some SED-derived properties are interdependent due to well-known degeneracies, such as the age-metallicity-attenuation relation (\citealp{2001ApJ...559..620P, 2022ApJ...927..170T}). We address this issue in more detail in Appendix \ref{degeneracy}.
Despite this limitation, this analysis provides valuable insights and helps us to obtain a broader understanding of the physics governing early galaxies.  

\begin{figure*}
\centering
\includegraphics[width=0.45\hsize]{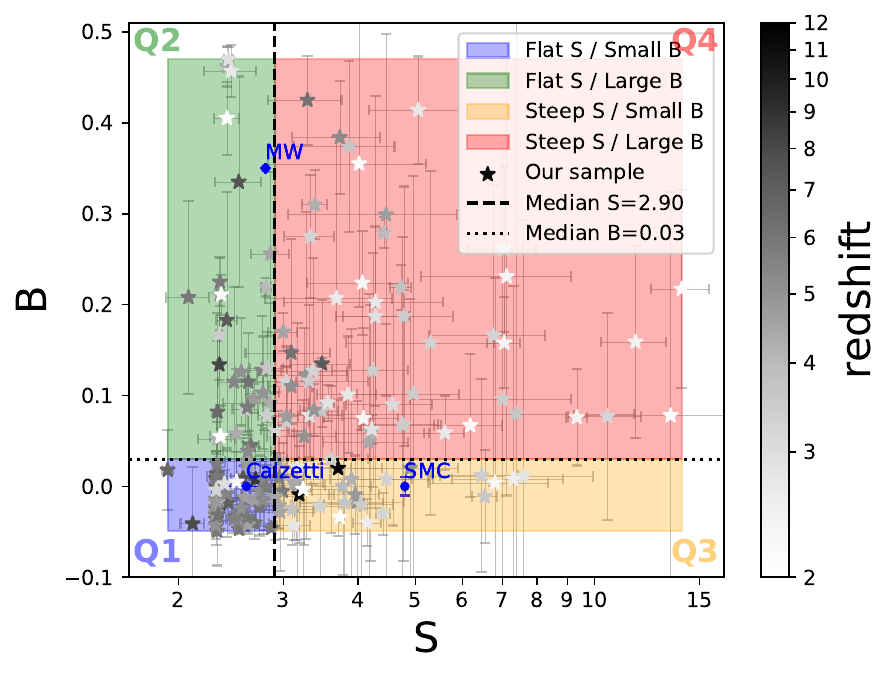}
\includegraphics[width=0.52\hsize]{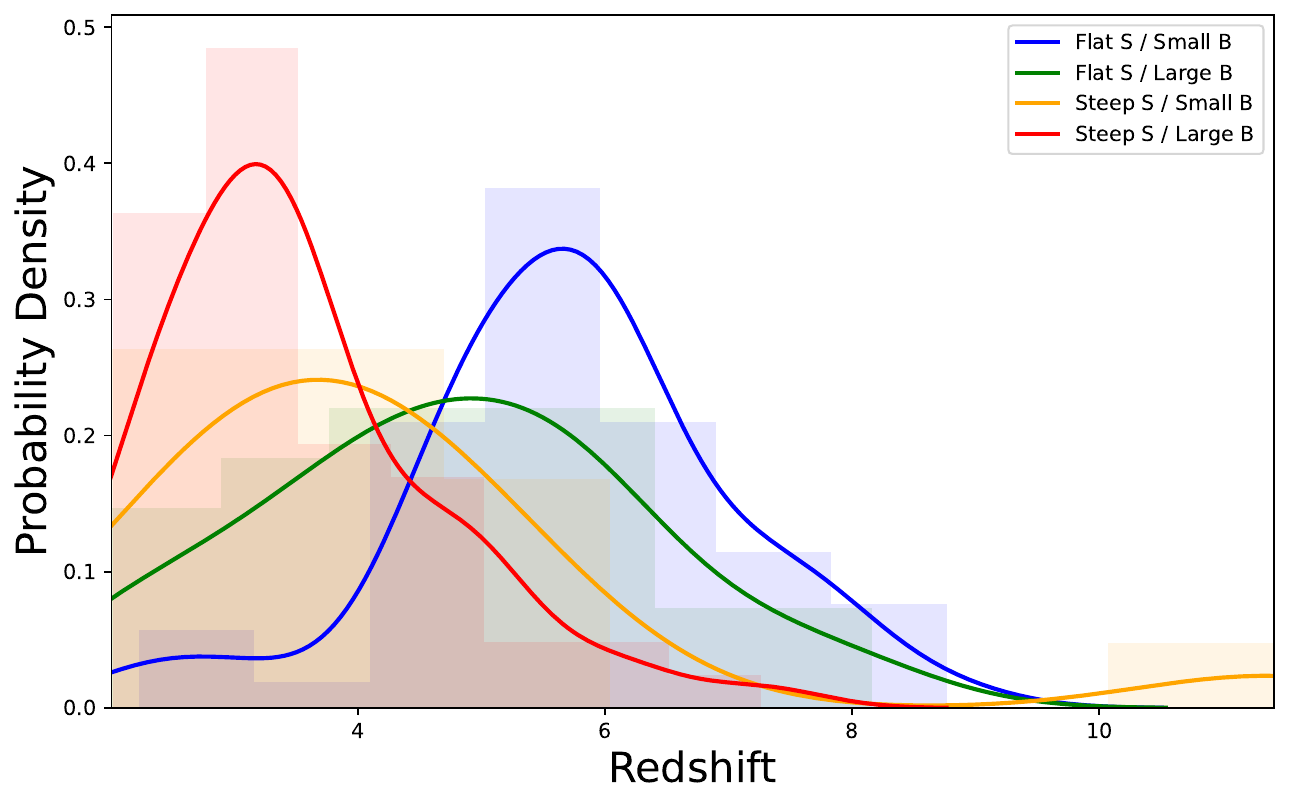}
\caption{Redshift evolution of the slope ($S$) and the UV bump ($B$). Left panel: $S$ and $B$, color-coded by redshift (grayscale), for our full galaxy sample. Stars represent individual measurements with $1\sigma$ uncertainties derived through a bootstrapping method. 
The $S-B$ parameter space is divided into four quadrants based on the median values of $S$ and $B$ values, indicated by dashed and dotted black lines, respectively. Sources with a flat $S$ and small $B$ (Calzetti-like curve) fall in the blue quadrant (Q1), those with a flat $S$ and large $B$ (MW-like curve) are in the green quadrant (Q2), galaxies with a steep $S$ and small $B$ (SMC-like curve) are in the orange quadrant (Q3), and sources with a steep $S$ and large $B$ occupy the red quadrant (Q4).  Literature results for nearby galaxies (\citealp{2020ARA&A..58..529S})  are depicted as blue symbols. Right panel:  Probability density distribution of redshift for galaxies in each quadrant of the $S-B$ parameter space. The probability density is estimated using a normalized histogram (shaded; \citealp{Hunter:2007}) and is overlaid with a kernel density estimation (solid lines; \citealp{Waskom2021}), which applies a Gaussian smoothing function to the discrete redshift values.
\label{B_S}
}
\end{figure*}

\subsection{Slope ($S$) and bump ($B$) versus redshift} \label{curve_z_trend}

Fig. \ref{B_S} illustrates the redshift evolution of the slope ($S$) and the UV bump ($B$). The left panel presents $S$ and $B$ parameters, color-coded by redshift (grayscale), distributed in four quadrants (Q1-Q4), delimited by the median values of $S$ ($\tilde{S} \sim 2.9$) and $B$ ($\tilde{B} \sim 0.03$) for the full galaxy sample. The right panel shows the redshift probability distribution functions for sources in each quadrant. 

At high redshift ($z > 5$), galaxy attenuation curves are typically Calzetti-like, that is, they are characterized by flat slopes and the absence of a UV bump. As a result, these galaxies predominantly occupy the blue quadrant (Q1). This trend can be explained if, at these redshifts, dust is primarily composed of large grains (produced in the core-collapse Type II supernovae (SNII) ejecta), leading to intrinsically flat and bump-free attenuation curves (\citealp{2022MNRAS.517.2076M, 2024NatAs.tmp...20M}). 

At lower redshifts ($z \sim 2-5$), galaxies exhibit a larger diversity in attenuation curve shapes, and their distributions are therefore spread across the remaining three quadrants. The migration of low-$z$ sources in the high-$S$ and high-$B$ parameter space reflects an evolution in the intrinsic dust properties that favors smaller carbonaceous dust grains. This transition can be driven by a combination of physical processes: (i) SNII-type dust is reprocessed in the ISM (e.g., \citealp{2018MNRAS.478.2851M, 2020MNRAS.492.3779H}), a process that converts large grains into small grains through both grain-grain collisions \citep[i.e., shattering,][]{Jones1996ApJ, Yan2004} and gas-grains collisions \citep[i.e., sputtering,][]{Draine:1979, Tielens:1994, BianchiSchneider2007}; (ii) the dominant dust production mechanism moves toward asymptotic giant branch (AGB) stars (\citealp{2006A&A...447..553F}) at later epochs.

In addition to redshift, $S$ and $B$ are influenced by $A_V$, which is a proxy for RT effects. It can shift the position of a galaxy in the $S$–$B$ parameter space in all directions. For instance, increasing $A_V$ would systematically shift sources within the blue quadrant. Additionally, $S$ depends on ${\langle a \rangle}_*^{\rm{m}}$ and sSFR, such that an aging stellar population and declining burstiness can drive a rightward shift in the $B$ versus $S$ plot. On the other hand, the $B$ parameter depends on $\rm{12+ \log{O/H}}$, meaning that an increasing metallicity can induce an upward shift in the $B$ versus $S$ plot.

\begin{figure*}
\centering
\includegraphics[width=0.49\hsize]{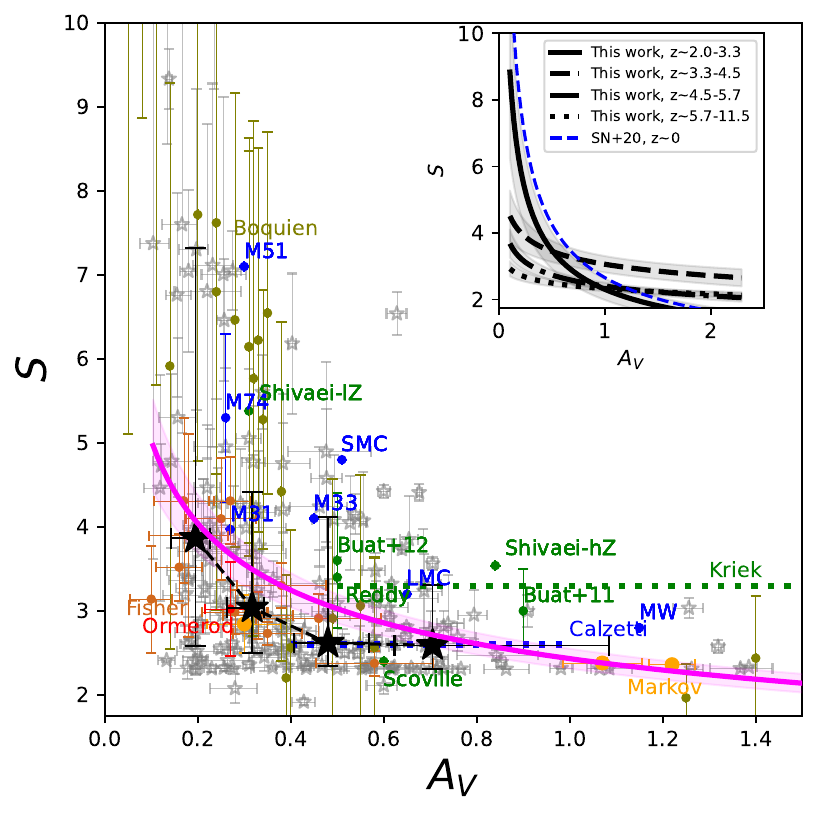}
\includegraphics[width=0.49\hsize]{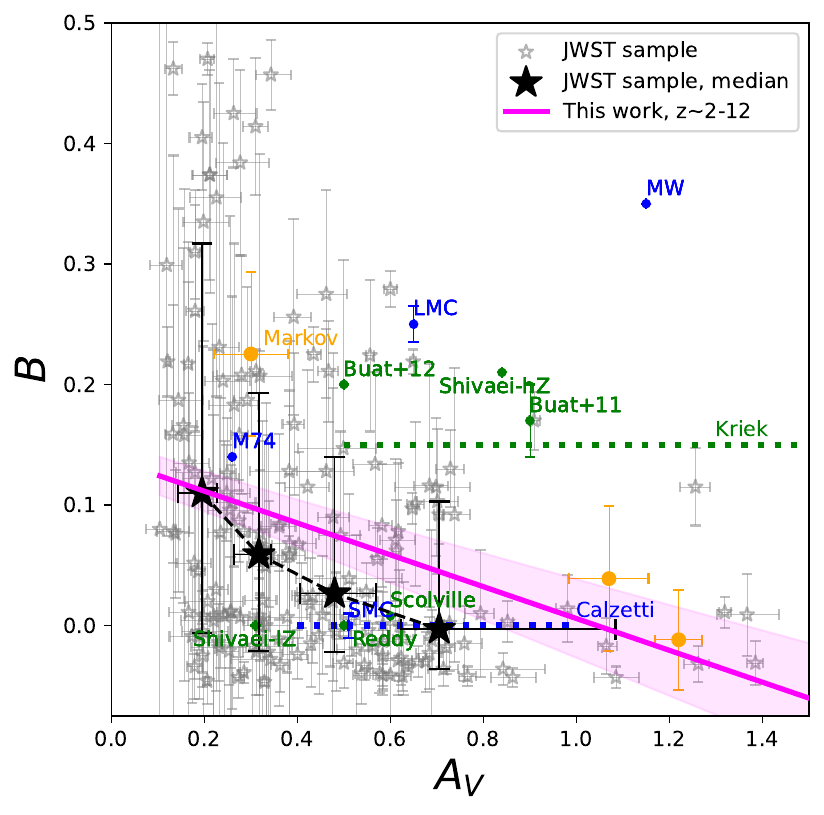}
\caption{Dust attenuation parameters as a function of $V$-band attenuation, $A_V$. The left panel shows the UV-optical slope (S), and the right panel displays the UV bump strength (B) as a function of $A_V$.  
Gray stars represent individual measurements of $S$ and $B$ for galaxies of our sample, with corresponding $1\sigma$ error bars derived through a bootstrapping method. Black stars indicate the median values of $S$ and $B$ for galaxies binned by $A_V$, with errors representing the $1\sigma$ dispersion of each bin. The best-fit correlations and their $1\sigma$ uncertainties for the entire sample and the intermediate- and high-$z$ subsets are shown by solid magenta lines, respectively. The shaded regions indicate the uncertainty ranges.  Literature results for nearby (\citealp{2019MNRAS.486..743D, 2020ARA&A..58..529S}) and intermediate-$z$ galaxies (\citealp{2011A&A...533A..93B, 2012A&A...545A.141B, 2015ApJ...800..108S, 2015ApJ...806..259R, 2020ApJ...899..117S}) are depicted as blue and green symbols, respectively. High-$z$ objects (\citealp{2022A&A...663A..50B, 2023A&A...679A..12M, 2025MNRAS.539..109F, ormerod2025detection2175aauvbump}) are shown as olive, orange, brown, and red symbols. The error bars indicate their $1\sigma$ uncertainties. The dotted blue and green lines indicate the average attenuation curve slopes and UV bump strengths for local starbursts with $A_V \sim 0.4-1.0$ (\citealp{2000ApJ...533..682C}) and for galaxies at $z \sim 0.5-2.0$ with $A_V \sim 0.5-2.25$ (\citealp{2013ApJ...775L..16K}), respectively. Inset panel: Black lines represent the best-fit correlations and their $1\sigma$ uncertainties for the subsets grouped by redshift. The shaded regions indicate the uncertainty ranges. The $S-A_V$ correlation from \cite{2020ARA&A..58..529S} for a nearby ($z \sim 0$) sample is shown as a dashed blue line in the left panel.  
\label{Av_SB}
}
\end{figure*}

\subsection{Slope ($S$) and bump ($B$) versus $A_V$} \label{curve_Av_trend}

In the correlation matrix (Fig. \ref{PCA}), we identified moderate $S-A_V$ and $B-A_V$  anticorrelations, suggesting that galaxies with low (high) $A_V$ tend to have steeper (flatter) slopes $S$ and stronger (weaker) UV bumps $B$. These trends are further illustrated in Fig. \ref{Av_SB}. Because a power-law correlation between $S$ and $A_V$ was reported in previous studies (e.g., \citealp{2020ARA&A..58..529S}), we modeled the $S-A_V$ correlation with a power-law fit, and we estimated the best-fit parameters using the \texttt{numpy polyfit} function
\begin{equation}
\log{S} = k_s \log{A_V} + n_s. \label{eq1}
\end{equation}%
The resulting best-fit model is reported in the left panel of Fig. \ref{Av_SB}.

The $B-A_V$ relation also exhibits a power-law behavior, which is consistent with predictions from RT models (\citealp{2004ApJ...617.1022P, 2016ApJ...833..201S}). 
The prior limits imposed on the UV bump parameter $c_4$ in the SED fitting procedure (see \citealp{2023A&A...679A..12M} for more details) mean that $c_4$, and consequently, the derived $B$ parameter, might occasionally take on slightly negative (unphysical) values, however. For this reason, we adopted a simpler linear fit to model the $B-A_V$ relation,
\begin{equation}\label{eqB}
B = k_b A_V + n_b,
\end{equation}%
and we represent the resulting best-fit model in the right panel of Fig. \ref{Av_SB}.
The best-fit coefficients of the power-law model, $k_s$, $n_s$, and the linear model, $k_b$, and $n_b$, for the full sample and redshift subsamples are instead reported in Table \ref{params}. 
We underline that in our sample, the $S-A_V$ and $B-A_V$ trends are independent of the remaining key drivers in attenuation curve variations, such as redshift, ${\langle a \rangle}_*^{\rm{m}}$, and sSFR (see Appendix \ref{App_AV} for further details).

As previously discussed in Sect. \ref{intro}, the $S-A_V$ power-law correlation was observed at lower redshifts ($z \lesssim 3$; e.g., \citealp{2020ARA&A..58..529S, 2020ApJ...888..108B}) and agrees with predictions from RT models (e.g., \citealp{2005MNRAS.359..171I}). Notably, \cite{2020ARA&A..58..529S} reported a steeper but qualitatively similar $S-A_V$ correlation for a sample of 23,000 galaxies at $ z\sim0$. 
We show the $S-A_V$ power-law correlations resulting from our analysis at different redshifts in the inset of Fig. \ref{Av_SB}, where we also report the $z\sim 0$ relation. This figure demonstrates a qualitative universality of the $S-A_V$ relation that smoothly evolves with redshift: The $S-A_V$ relation is steeper at low redshifts ($k_s = -0.68$ at $z \sim 0$, \citealp{2020ARA&A..58..529S}) and becomes progressively flatter at intermediate ($k_s \sim -0.59$ at $z \sim 2-3.3$), and at increasingly higher redshifts ($k_s \sim -0.18$ at $z \sim 3.3-5.7$ and $k_s \sim -0.10$ at $z \sim 5.7-11.5$, respectively). This is consistent with the findings of \citet[][see Fig. 4a of their work]{2024NatAs.tmp...20M}.

In the high-$A_V$ limit ($A_V \gtrsim 1$), average attenuation curves tend to be Calzetti-like, exhibiting predominantly flat slopes ($S \lesssim 3$) and negligible bumps ($B \lesssim 0.1$). This behavior likely reflects strong RT effects at high $A_V$ (\citealp{1984ApJ...287..228N, 2018ApJ...869...70N}).
Conversely, in the low-$A_V$ regime ($A_V \lesssim 0.5$), where RT effects are not expected to dominate (\citealp{1984ApJ...287..228N, 1994ApJ...429..582C}), the attenuation curves are typically steeper ($S \gtrsim 3$), with more prominent bumps ($B \gtrsim 0.1$; Fig. \ref{Av_SB}).
Additionally, in the low-$A_V$ regime, attenuation curves exhibit greater dispersion, which might reflect secondary dependences on other key parameters, such as $z$, ${\langle a \rangle}_*^{\rm{m}}$, and sSFR.

Finally, the combined $S-A_V$ and $B-A_V$ relations enabled us to constrain the average attenuation curve of a galaxy based on a single parameter ($A_V$) within a specific range of redshift or stellar age (or, equivalently, sSFR). 

\begin{figure}
\centering
\includegraphics[width=\hsize]{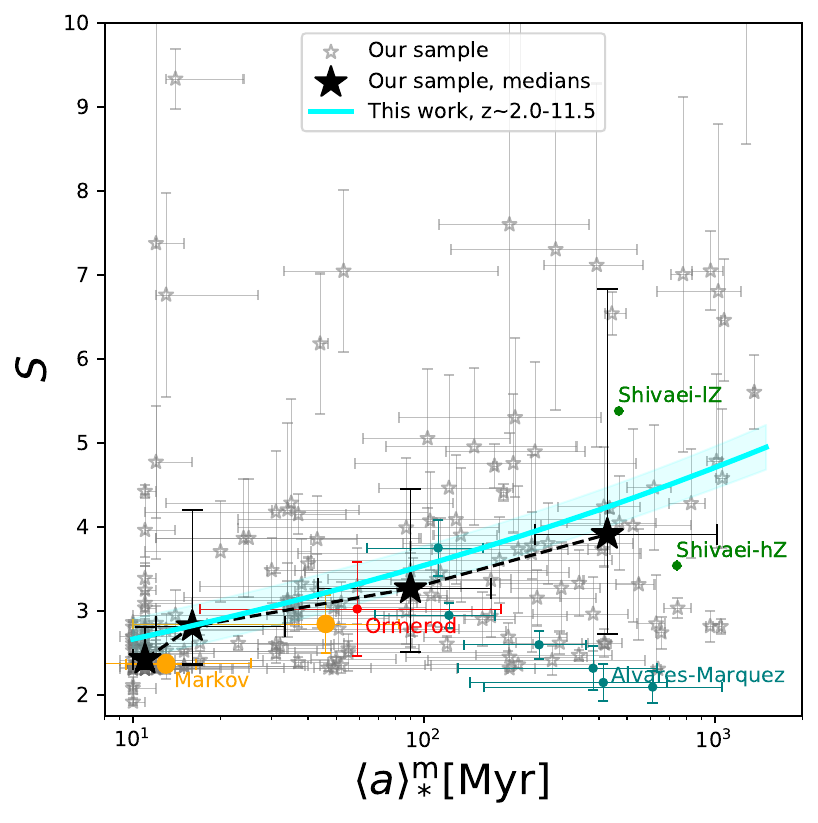}
\caption{UV-optical slope (S) as a function of the mass-weighted stellar age (${\langle a \rangle}_*^{\rm{m}}$.   
Gray stars represent individual $S$ measurements from our sample, with $1\sigma$ uncertainties derived through a bootstrapping method. Black stars indicate the median $S$ values for galaxies grouped by ${\langle a\rangle}_*^{\rm{m}}$, and the error bars show the $1\sigma$ dispersion within each bin. The solid cyan line represents the best-fit correlation for the entire sample using a power-law model (Eq. \ref{eq1}), and the shaded regions indicate the uncertainty ranges. Literature results for intermediate-$z$ sources or stacks are illustrated as green (\citealp{2020ApJ...899..117S, 2022MNRAS.513.4431B})  and teal (\citealp{2019A&A...630A.153A}) symbols. Finally, high-$z$ objects (\citealp{2023A&A...679A..12M, ormerod2025detection2175aauvbump}) are depicted as orange and red symbols, respectively. The error bars represent their $1\sigma$ uncertainties. 
\label{age_S}
}
\end{figure}

\subsection{Slope ($S$) versus stellar age ${\langle a \rangle}_*^{\rm{m}}$}  \label{S_age_trend}

Our analysis revealed a positive correlation between $S$ and ${\langle a \rangle}_*^{\rm{m}}$ ($r = 0.47$), but we found no significant correlations between ${\langle a \rangle}_*^{\rm{m}}$ and $B$ ($r = 0.077$). 
We report the trend of increasing $S$ with ${\langle a \rangle}_*^{\rm{m}}$ in Fig. \ref{age_S}. We fit the $S-{\langle a \rangle}_*^{\rm{m}}$ correlation with a power law (see Eq. \ref{eq1}, and Fig. \ref{age_S}), and we summarize the best-fit coefficients in Table \ref{params}.

We examined the consistency of the $S-{\langle a \rangle}_*^{\rm{m}}$ correlation in the galaxy subsets grouped by redshift, $A_V$, and sSFR (see Appendix \ref{App_age}). 
The right panel of Fig. \ref{S_age_AvsSFR} shows that the $S-{\langle a \rangle}_*^{\rm{m}}$ trends become statistically insignificant when we controlled for sSFR. In addition, our Pearson correlation analysis revealed a strong anticorrelation between ${\langle a \rangle}_*^{\rm{m}}$ and $\rm{sSFR}$ ($r = -0.54$), consistent with the interpretation that the sSFR serves as a proxy for the mean stellar age (e.g., \citealp{2024arXiv241014671L}). This suggests that the $S-{\langle a \rangle}_*^{\rm{m}}$ trend partially reflects the $S-{\rm{sSFR}}$ anticorrelation, and vice versa. 

Similarly, the $S-{\langle a \rangle}_*^{\rm{m}}$ trend becomes somewhat weaker when galaxies are binned by redshift (the left panel of Fig. \ref{S_age_AvsSFR}). These results, along with the inverse relation between the ${\langle a \rangle}_*^{\rm{m}}$ and redshift ($r = -0.37$; Fig. \ref{PCA}), suggest that the $S-{\langle a \rangle}_*^{\rm{m}}$ correlation in part depends on the previously reported $S-z$ trend (\citealp{2024NatAs.tmp...20M}).

To further analyze the redshift dependence of the $S-{\langle a \rangle}_*^{\rm{m}}$ trend, we show in the left panel of Fig. \ref{S_reds_age} the redshift evolution of $S$. We color-code individual sources with their ${\langle a \rangle}_*^{\rm{m}}$. This figure shows that young stellar populations (${\langle a \rangle}_*^{\rm{m}} \lesssim 50$ Myr) are predominantly found at high redshift ($z \gtrsim 5$) when the attenuation curves are flattest ($S \lesssim 3$), while older sources (${\langle a \rangle}_*^{\rm{m}} \gtrsim 100$ Myr) are typically observed at lower redshifts ($z \lesssim 4$), when the attenuation curves become steeper ($S \gtrsim 4$).

\begin{figure}
\centering
\includegraphics[width=\hsize]{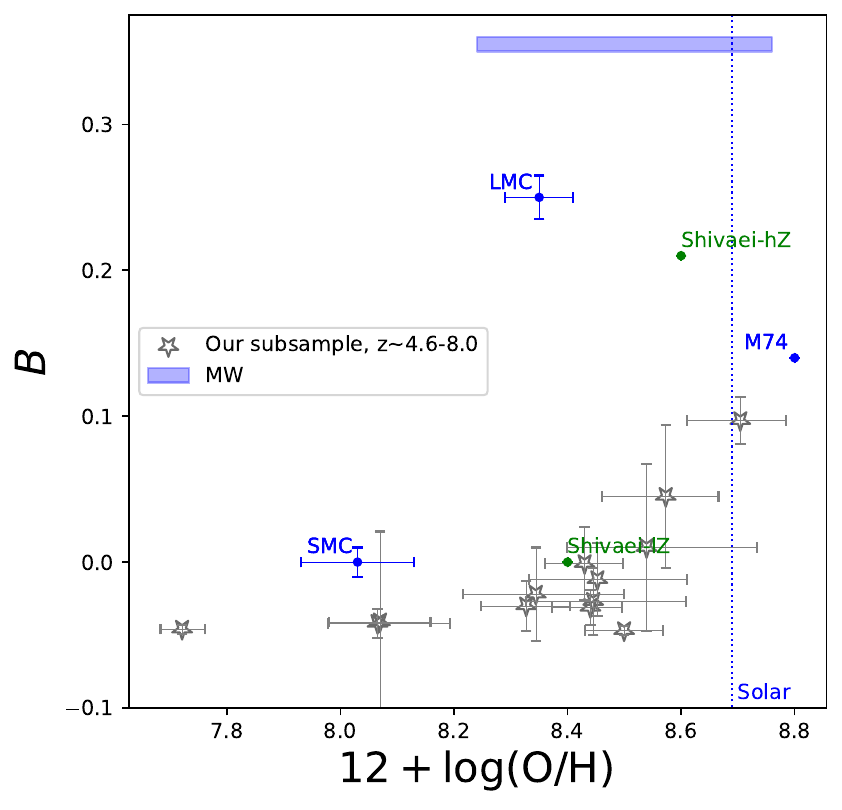}
\caption{UV bump (B) as a function of the oxygen abundance ($12 + \log(\rm{O/H})$).
Stars represent individual measurements from our subsample of sources with oxygen abundance measurements, with $1\sigma$ uncertainties derived through a bootstrapping method.  For reference, the solar oxygen abundance is $\rm{{12 + \log(O/H)}_{\odot} = 8.69}$ (dotted blue line; \citealp{2001ApJ...556L..63A}). The MW oxygen abundance is in the $\rm{{12 + \log(O/H)}\sim 8.24-8.76}$ range, depending on the Galactic radius (horizontal blue strip; \citealp{2023A&A...676A..57P}).   Literature results for low-$z$ (\citealp{1992ApJ...384..508R, 2019A&A...621A..51H, 2019MNRAS.486..743D}) and intermediate-$z$ (\citealp{2020ApJ...899..117S}) sources are shown as blue and green symbols, respectively. 
\label{B_OH}
}
\end{figure}

\subsection{UV Bumb ($B$) versus oxygen abundance ($12 + \log(\rm{O/H})$)}  \label{S_Z_trend}

We further investigated the correlations between the galaxy properties and restricted our sample to the subset of 13 galaxies ($\sim 8\%$ of our full sample) with oxygen abundance measurements (see Sect. \ref{method_metallicity}).

We found a strong correlation between the oxygen abundance ($12 + \log(\rm{O/H})$ and the UV bump strength ($r = 0.66$ for $B$ and $r = 0.65$ for $c_4$),  
but no significant correlation was found between the slope ($S$) and the oxygen abundance ($r = 0.15$).  
In Fig. \ref{B_OH} we show the $B-12 + \log(\rm{O/H})$ relation. The large uncertainties associated with the small size of this sample prevent us from determining a converging best-fit model for this trend.  
In Sect. \ref{origin_OH} we discuss the possible underlying origin of the $B-12 + \log(\rm{O/H})$ trend.

While these results provide useful insights, they should be interpreted with caution because the number of sources was limited. A larger galaxy sample with oxygen abundance measurements is necessary to confirm the correlations we identified in our analysis and to ensure statistical robustness.

\section{Trends between fundamental galaxy properties} \label{global_trends}

We used the correlation matrix (Fig. \ref{PCA}) to identify the general trends among the global properties derived from the SED fitting (redshift, $A_V$, SFR $\log{M_*}$, $\rm{sSFR}$, ${\langle a \rangle}_*^{\rm{m}}$, and $\log{U}$) for the full galaxy sample. Furthermore, we considered the same correlations with the oxygen abundance as were derived using the direct $T_e$ method for our subset of sources with bright emission lines. 
The emerging picture is summarized below.  

\begin{description}
\item[A) $A_V$:] Dustier galaxies generally form more stars ($r = 0.6$) and are more massive ($r = 0.33$; Fig. \ref{mass_Av}).
Additionally, $A_V$ is anticorrelated with the mean stellar age ${\langle a \rangle}_*^{\rm{m}}$ when $\log{M_*}$ is fixed, which indicates that galaxies with a higher dust attenuation typically host younger stellar populations. 
\begin{figure}
\centering
\includegraphics[width=\hsize]{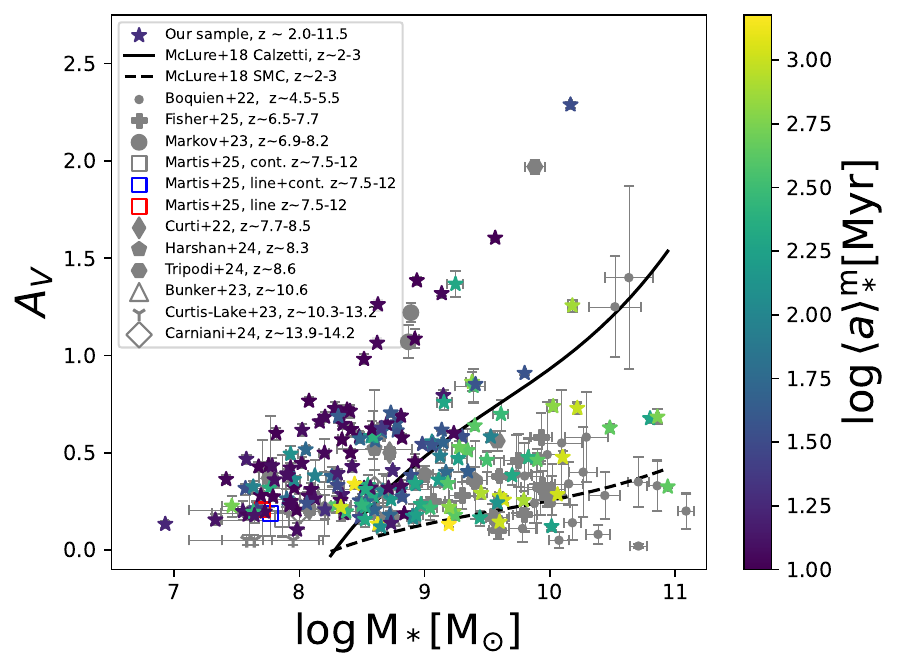}
\caption{$V$-band attenuation ($A_V$) vs. stellar mass ($\log{M}$) color-coded by the mass-weighted stellar age (${\langle a \rangle}_*^{\rm{m}}$). Our sources are represented as stars. The solid and dashed black lines indicate the predicted $A_V-\log{M}$ relation by  \cite{2018MNRAS.476.3991M}, based on the assumption of Calzetti and SMC-like attenuation laws, respectively.  We also include literature results for high-$z$ sources (\citealp{2022A&A...663A..50B, 2023NatAs.tmp...66C, 2023MNRAS.518..425C, 2023A&A...677A..88B, 2023A&A...679A..12M, 2024ApJ...977L..36H, 2024Natur.633..318C, 2024arXiv241204983T, ormerod2025detection2175aauvbump}), shown as gray symbols. Median values for continuum, continuum+line,
and line-only samples of $z\sim7.5-12$ galaxies from
the JWST Technicolor (Program ID 3362) and CANUCS (\citealp{Willott_2022}) surveys are indicated as empty gray, blue, and red squares, respectively (\citealp{martis2025canucstechnicolorjwstmediumband}). The $1\sigma$ uncertainties are indicated as error bars.
\label{mass_Av}
}
\end{figure} 
\item[B) ${\langle a \rangle}_*^{\rm{m}}$ and sSFR:] Young stellar populations are predominantly found in star-bursting galaxies ($r = -0.54$) at high redshift ($r = -0.37$), as shown in Fig. \ref{age_z}. 
Additionally, young starbursts tend to be metal-poor ($r = 0.44$) and low-mass sources ($r = 0.44$). 
\begin{figure}
\centering
\includegraphics[width=\hsize]{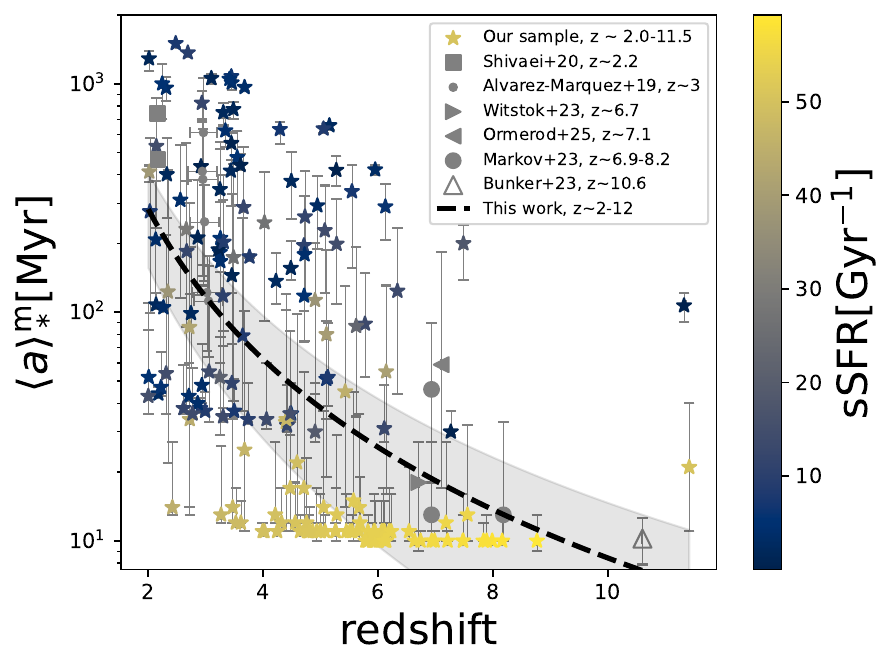}
\caption{Mass-weighted stellar age (${\langle a \rangle}_*^{\rm{m}}$) as a function of redshift ($z$), color-coded by the sSFR. Stars represent individual measurements for our sample, and the error bars represent the $1\sigma$ uncertainties. The dashed black line represents the best-fit power-law model for the entire galaxy sample. Comparative literature results for intermediate-$z$ (\citealp{2019A&A...630A.153A, 2020ApJ...899..117S}) and high-$z$ objects (\citealp{2023Natur.621..267W, 2023A&A...677A..88B, 2023A&A...679A..12M, ormerod2025detection2175aauvbump}) are depicted as gray symbols, with their respective $1\sigma$ uncertainties.
\label{age_z}
}
\end{figure} 
\item[C) $M_*$, SFR, metallicity, $\log{U}$:] Massive sources have a higher metallicity ($r = 0.64$), that is, oxygen abundance ($r = 0.63$), 
and increased star formation ($r = 0.44$), as shown in Fig. \ref{mass-Z}. This follows the well-known mass-metallicity relation (e.g., \citealp{2004ApJ...613..898T}). Furthermore, massive sources are generally characterized by higher $A_V$ ($r = 0.33$) and a more evolved stellar population ($r = 0.44$), as shown in Fig. \ref{mass_Av}. Finally, massive sources also exhibit weaker ionization $\log{U}$ ($r = -0.48$), and by definition, they have a lower sSFR ($r = -0.62$). 
\begin{figure}
\centering
\includegraphics[width=\hsize]{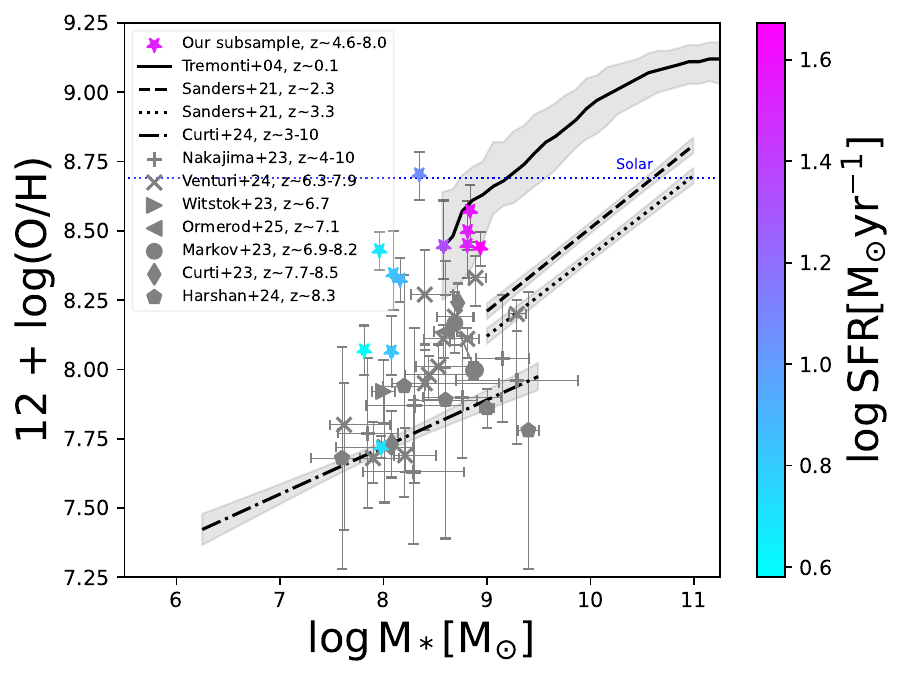}
\caption{ Mass-metallicity relation: Oxygen abundance ($12 + \log(\rm{O/H})$) vs. stellar mass ($\log{M}$) color-coded by the SFR. $12 + \log(\rm{O/H})$ is estimated using direct $T_e$ calibrations (\citealp{2024ApJ...962...24S}) for a subset of 13 sources ($\sim 8\%$) with bright emission lines (see Sect. \ref{S_Z_trend} for details). Our galaxies are represented as inverted stars. The mass-metallicity relations found for nearby (\citealp{2004ApJ...613..898T}, intermediate-$z$ (\citealp{2021ApJ...914...19S}), and high-$z$ (\citealp{2024A&A...684A..75C}) objects are depicted as solid, dashed and dotted, and dashed-dotted black lines, respectively. We also show the literature results for high-$z$ sources (\citealp{2023ApJS..269...33N,  2023MNRAS.518..425C, 2023Natur.621..267W, 2023A&A...679A..12M, 2024ApJ...977L..36H, 2024A&A...691A..19V, ormerod2025detection2175aauvbump}).
\label{mass-Z}
}
\end{figure}
\end{description}

\section{Discussion} \label{Discussion}

In this section, we connect the observed trends between dust attenuation properties and fundamental galaxy properties (Sec. \ref{Results}) with the correlations found among galaxy properties (Sect. \ref{global_trends}). The final goal is to identify the key physical mechanisms that shape dust attenuation curves through cosmic time in the broader context of galaxy formation and evolution. 

The origin of the $S-z$ and $B-z$ trends was discussed in detail by \cite{2024NatAs.tmp...20M} and the $S-A_V$ and $B-A_V$ trends agree with predictions from RT calculations (\citealp{2018ApJ...869...70N}). We therefore focus in this section on the $S-{\langle a \rangle}_*^{\rm{m}}$ (or $S-\rm{sSFR}$) and $B-12 + \log(\rm{O/H})$ trends. 

Although there is no clear consensus in the literature, several physical mechanisms have been proposed as drivers of these correlations. We list them below.
\begin{description}
\item[M1:] {Additional attenuation in stellar birth clouds}. Starburst galaxies hosting young stars embedded in short-lived molecular birth clouds (${\langle a \rangle}_* \lesssim 10 \ \rm{Myr}$; \citealp{2000ApJ...539..718C}) experience additional attenuation along the LOS beyond the attenuation that is contributed by the diffuse ISM  (\citealp{2000ApJ...542..710G, 2007MNRAS.375..640P}). In the optically thick (high-$A_V$) limit, the flattening of the attenuation curve in the UV wavelength range is driven by radiation from unobscured stars and scattering of UV light into the LOS (\citealp{2018ApJ...869...70N}). This scenario agrees to some extent with the global trend A) discussed in Sect. \ref{global_trends}.
\item[M2:] {Complex dust-star geometry}. The increasing complexity of the dust-star geometry in galaxies results in flatter attenuation curves (\citealp{1996ApJ...463..681W, 2018ApJ...869...70N}). In particular, \cite{2018ApJ...869...70N} emphasized that the spatial distribution of dust relative to stellar populations of different stellar ages is the key factor that shapes the attenuation curve. Specifically, a higher fraction of unobscured young stars flattens the attenuation curve by making the galaxy more transparent to UV photons. 
\item[M3:] {Strong radiation field in starbursts}. Intense radiation fields in young highly star-forming systems can destroy smaller dust grains (the carriers of the steep slopes), shift the grain size distribution toward larger grains, and flatten the attenuation curves (\citealp{2003ApJ...594..279G, 2011A&A...533A.117F, 2013ApJ...775L..16K, 2020ApJ...892...84T}). This effect is consistent with the global trend B) described in Sect. \ref{global_trends}. Conversely, some studies suggest that strong radiation fields can also lead to the formation of ultrasmall dust grains, including PAHs, which are commonly considered to be the carriers of the UV bump feature (\citealp{1990A&A...237..215D, 1992A&A...259..614S, 2005Sci...307..244B}) through increased shattering and UV-driven aromatization (\citealp{2023ApJ...951..100N, ormerod2025detection2175aauvbump}).
\item[M4:] {Variation in intrinsic dust properties with $z$}. A shift in the dust grain size distribution toward larger grains in high-$z$ sources results in flatter curves  (\citealp{2018MNRAS.478.2851M, 2020MNRAS.492.3779H, 2022MNRAS.517.2076M, 2024NatAs.tmp...20M}). This shift reflects changes in the dominant dust production mechanism at early cosmic epochs,  where SNII ejecta play a primary role in the dust formation (\citealp{2001MNRAS.325..726T, 2007ApJ...666..955N, 2014Natur.511..326G, 2024arXiv241014671L, 2025ApJ...985L..21M}).
\end{description}

To summarize, M1 and M2 are closely associated with RT effects. At the same time, M3 and M4 involve variations in dust properties: In the case of M3, variations in the dust properties are linked to dust re-processing, while in the case of M4, they are caused by changes in the intrinsic dust properties associated with different dust producers. Keeping in mind these mechanisms, we now determine the primary drivers of the $S-{\langle a \rangle}_*^{\rm{m}}$ and $B-12 + \log(\rm{O/H})$ relations.

\subsection{Origin of the slope $S$ - stellar age correlation} \label{origin_age}

Our results shows a moderate correlation (r = 0.47) between the slope ($S$) and the stellar age (${\langle a \rangle}_*^{\rm{m}}$) and an anticorrelation between $S$ and sSFR ($r = -0.38$), such that $S$ flattens with decreasing stellar age (${\langle a \rangle}_*^{\rm{m}}$) and increasing sSFR (Figs. \ref{age_S} and \ref{sSFR_S}, respectively). 
The strong anticorrelation between ${\langle a \rangle}_*^{\rm{m}}$ and sSFR (r = -0.54; Fig. \ref{PCA}) further highlights their close connection, with the inverse of sSFR serving as a proxy for mean stellar age. Additionally, a lack of statistically significant $S-{\langle a \rangle}_*^{\rm{m}}$ trends is observed after fixing for sSFR, except in one bin (Fig. \ref{S_age_AvsSFR}, right panel).
These results strongly suggest that M3 is one of the primary factors driving the $S-{\langle a \rangle}_*^{\rm{m}}$ trend.
According to M3, the flattening of the slope (S) is associated with the destruction of small dust grains in strong radiation fields associated with young starbursts (e.g., \citealp{2013ApJ...775L..16K, 2025PASA...42...22B}), particularly at high-$z$, where the ISM conditions are more extreme (\citealp{2019MNRAS.489....1F, 2020MNRAS.495L..22V, 2021MNRAS.505.5543V, 2022A&A...663A.172M}). 

Next, a moderate anticorrelation between ${\langle a \rangle}_*^{\rm{m}}$ and redshift ($r = -0.37$, Fig. \ref{PCA}), along with the fact that $S-{\langle a \rangle}_*^{\rm{m}}$ trend is less pronounced after fixing the redshift (Fig. \ref{S_age_AvsSFR}, left panel), suggest that M4 likely contributes to the observed $S-{\langle a \rangle}_*^{\rm{m}}$ correlation. M4 involves the flattening of $S$ with decreasing ${\langle a \rangle}_*^{\rm{m}}$, which might reflect the underlying $S-z$  trend discussed by \cite{2024NatAs.tmp...20M}. 

The lack of significant correlations between  ${\langle a \rangle}_*^{\rm{m}}$ and $A_V$ combined with the consistency of the $S-{\langle a \rangle}_*^{\rm{m}}$ relation in all $A_V$ bins (Fig. \ref{S_age_AvsSFR}, middle panel) suggests that M1 alone cannot account for the observed $S-{\langle a \rangle}_*^{\rm{m}}$ trend.

Finally, M2 (associated with the increasing complexity of dust-star geometry, i.e., an increasing fraction of unobscured young stars) is challenging to test observationally and relies on theoretical works involving cosmological hydrodynamic simulations (e.g., \citealp{2018ApJ...869...70N}).
Within the redshift range of our sample ($z\sim2-11.5$), the galaxy morphologies are overall expected to be clumpy with strong dust-gas-star decoupling   (\citealp{2018MNRAS.477..552B, 2018MNRAS.478.1170C, 2024arXiv240218543F}).  
In a typical young clumpy galaxy within our sample (i.e., with lower ${\langle a \rangle}_*^{\rm{m}}$), an important fraction of young blue stars are expected to be unobscured. The UV radiation from unobscured young stars, combined with the scattering of UV light into the line of sight in clumpy geometries, results in a flatter attenuation curve slope ($S$; \citealp{2018ApJ...869...70N}).
Therefore, this mechanism is consistent with the observed $S-{\langle a \rangle}_*^{\rm{m}}$ anticorrelation and cannot be excluded.

In summary, the observed $S-{\langle a \rangle}_*^{\rm{m}}$ correlation, characterized by the flattening of $S$ with decreasing ${\langle a \rangle}_*^{\rm{m}}$, appears to be primarily driven by the starburst nature of these sources. This reflects a shift in reprocessed dust properties through the destruction of small dust grains in the intense radiation fields of dusty starburst galaxies (M3). Contributions from the complex dust-star geometry (M2) and evolving intrinsic dust properties with redshift (M4) are likely secondary effects because they partially overlap with the trends associated with sSFR.  

\subsection{Origin of the UV bump ($B$)- oxygen abundance correlation} \label{origin_OH}

Our analysis showed a strong correlation ($r = 0.66$) between the UV bump parameter ($B$) and the oxygen abundance ($12 + \log(\rm{O/H})$; Fig. \ref{B_OH}). Furthermore, we found that the oxygen abundance is anticorrelated with redshift ($r = -0.4$) and is correlated with stellar age ($r = 0.33$) and stellar mass ($r = 0.63$). The combination of these results suggests that the formation of small carbonaceous dust grains or PAHs is more favorable in more evolved massive metal-rich galaxies at lower redshifts \citep[consistently with the findings by][]{2012A&A...545A.141B, 2015ApJ...814..162Z, 2020ApJ...899..117S}. In particular, the possible PAH-age correlation may reflect the delayed injection of carbon dust into the ISM by AGB stars \citep{Galliano2008}, the primary sources of PAHs and carbon dust in galaxies. This scenario supports mechanism M4 as an important driver of the $B-12 + \log(\rm{O/H})$ trend. 

We note, however, that although a positive $B$-metallicity correlation is both theoretically expected and observed in local extinction (e.g., \citealp{1989ApJ...345..245C, 2024ApJ...970...51G}) and attenuation curves (e.g., \citealp{2020ApJ...899..117S, 2022MNRAS.514.1886S}), evidence for this correlation is not universally found (e.g., \citealp{2018ApJ...859...11S, 2025PASA...42...22B}). This might be due to RT effects (see M1 and M2) that can considerably suppress the intrinsic UV bump in attenuation curves (e.g., \citealp{2000ApJ...528..799W, 2016ApJ...833..201S}).

\section{Summary and conclusions}

This paper extended the analysis of the  JWST NIRSpec prism spectroscopic data of 173 dusty high-redshift ($z\sim2-11.5$) galaxies presented by \cite{2024NatAs.tmp...20M}. Using our customized SED fitting approach (\citealp{2023A&A...679A..12M, 2024NatAs.tmp...20M}), we inferred a diverse range of dust attenuation
curves in that study and showed that their shape evolves with redshift. We also discussed potential origins of the observed trends. 

In this work, we used the same sample of galaxies to search for correlations between the parameters that govern the shape of the dust attenuation curves (i.e., UV-optical slope $S$ and UV bump parameter $B$) and the galaxy properties, both derived from the SED fitting (i.e., $V$-band attenuation $A_V$, stellar mass $\log{M_*}$, mass-weighted stellar age ${\langle a \rangle}_*^{\rm{m}}$, SFR, sSFR, metallicity $Z$, and ionization parameter $\log{U}$) and from optical emission lines (i.e., oxygen abundance $12 + \log(\rm{O/H})$). 

We identified the main trends in the correlations among dust parameters ($S$ and $B$) and galaxy properties as listed below.

\begin{itemize}
    \item Moderate correlations between $S-A_V$ and $B-A_V$, where galaxies with lower (higher) $A_V$ exhibit steeper (flatter) slopes $S$ and stronger (weaker) UV bumps $B$. These trends are overall independent of other primary factors driving variations in $S$ and $B$, such as redshift, stellar age (${\langle a \rangle}_*^{\rm{m}}$), and sSFR. Additionally, these trends are consistent across all redshifts and agree well with predictions from RT models.
    \item A moderate positive $S- {\langle a \rangle}_*^{\rm{m}}$ trend, where the slope ($S$) steepens as the galaxy stellar population ages. Additionally, we observed a moderate negative $S-\rm{sSFR}$  correlation that probably largely reflects the underlying $S- {\langle a \rangle}_*^{\rm{m}}$ relation caused by the strong inverse correlation between ${\langle a \rangle}_*^{\rm{m}}$ and sSFR. We hypothesized that the origin of the $S - {\langle a \rangle}_*^{\rm{m}}$ and  $S-\rm{sSFR}$ trends most probably lies in the starburst nature of these galaxies and that additional contributions are connected to the underlying redshift evolution of the intrinsic dust properties and dust-star geometry.   
    \item A strong positive correlation between the UV bump strength ($B$) and oxygen abundance ($12 + \log(\rm{O/H})$) for a subset of galaxies with oxygen abundance measurements. This trend is likely driven by intrinsic dust properties because extinction curves typically exhibit the same behavior. 
\end{itemize}
We identified the main trends for correlations among the galaxy properties as listed below.
\begin{itemize}
\item $A_V$: Dustier sources are typically more massive and form more stars.
\item ${\langle a \rangle}_*^{\rm{m}}$ and sSFR: Young starburst galaxies are typically found at higher redshifts and have lower metallicities and stellar masses. 
\item $\log{M_*}$, $\rm{SFR}$, $Z$ ($12 + \log(\rm{O/H})$), and $\log{U}$: Massive galaxies typically form more stars, are richer in metals, are more evolved, and are dustier. Their ionization and sSFR are lower.
\end{itemize}

Building on the analysis of \cite{2024NatAs.tmp...20M} and this work, we found that the shape of the dust attenuation curves is primarily correlated with four key galaxy properties: 1) redshift ($z$), which serves as a proxy for variations in the dust properties, both driven by different dust production mechanisms and by re-processing effects that occur in the ISM; 2) V-band attenuation ($A_V$), which reflects the impact of RT effects; 3) mass-weighted stellar age (${\langle a \rangle}_*^{\rm{m}}$) or sSFR, which trace the radiation field strength; and 4) oxygen abundance $12 + \log(\rm{O/H})$, which may be linked to intrinsic dust properties. 

The trends presented by \citet{2024NatAs.tmp...20M} and in this work collectively outline a broader picture of the dust lifecycle in galaxies. In the early Universe, galaxies typically had relatively flat featureless attenuation curves, primarily because of their intrinsic dust properties, which were dominated by large silicate-based grains that formed in SN ejecta. Over time, several processes contributed to the steepening of the attenuation curve and the emergence of the UV bump feature: 1) large grains are reprocessed in the ISM into smaller ones through shattering and sputtering; 2) AGB stars begin to dominate the dust production and contribute to smaller carbon-rich grains; and 3) the progressive metal enrichment of the ISM enhances the complexity of dust. Additional processes can instead lead to a flattening of the attenuation curve and suppression of the UV bump: 1) grain growth through accretion and coagulation in the ISM, which shifts the size distribution toward larger grains; and 2) intense radiation fields in young starburst galaxies, which can destroy small grains. Finally, the total amount of dust along the LOS as well as the spatial distribution of dust relative to stars can significantly alter the observed attenuation curve through RT effects that mask the underlying intrinsic extinction law.

Except for \cite{2024NatAs.tmp...20M} and this work, only a few empirical studies have explored these correlations in detail, particularly at high redshifts ($z \gtrsim 4$), where the connection between dust properties and fundamental galaxy properties remains largely unconstrained. Our results highlight the need for a comprehensive approach, where all key correlations between attenuation curves and evolving galaxy properties are jointly analyzed to separate the complex interplay between the underlying physical mechanisms. It is essential to apply the method we adopted in this work to larger galaxy samples to separate the physical drivers of attenuation curve variations. Furthermore, observations of point-like sources such as quasars and GRBs, which directly probe the intrinsic dust properties, are necessary (\citealp{2010A&A...523A..85G,2018A&A...609A..62B, 2018MNRAS.480..108Z}). Additionally, spatially resolved observations of galaxies are important for studying attenuation laws along different lines of sight and accounting for dust-star geometry effects (\citealp{2019MNRAS.486..743D, 2025PASA...42...22B, ormerod2025detection2175aauvbump}). Finally, cosmological hydrodynamical simulations incorporating dust evolution models are essential for constraining the role of RT effects in shaping attenuation curves (e.g., \citealp{2021MNRAS.506.3946D, 2024A&A...687A.240D}).

\begin{acknowledgements}
       We thank the anonymous referee for their valuable comments and suggestions, which helped improve the clarity and quality of this work.
       VM, MB, RT, and NM acknowledge support from the ERC Grant FIRSTLIGHT and the Slovenian National Research Agency
       ARRS through grants N1-0238, P1-0188.
       VM and AP acknowledge support from the ERC Advanced Grant INTERSTELLAR H2020/740120. 
       Any dissemination of results must indicate that it reflects only the author’s view and that the Commission is not responsible for any use that may be made of the information it contains.
       The data products presented herein were retrieved from the Dawn JWST Archive (DJA). DJA is an initiative of the Cosmic Dawn Center, which is funded by the Danish National Research Foundation under grant No. 140.
       We thank M. Decleir, I. Shivaei, and J. Witstok for sharing their findings on the properties of the attenuation curve. 
       We gratefully acknowledge the computational resources of the Center for High-Performance Computing (CHPC) at SNS.
\end{acknowledgements}

\bibliographystyle{aa} 
\bibliography{biblio}

\begin{appendix}  
\onecolumn
\section{Photometric bands}\label{App_photo}

In Table \ref{bands}, we report all photometric bands used to correct the NIRSpec prism spectra for slit losses. The photometry includes 16–30 filters, depending on the field, from HST/ACS, HST/WFC3, JWST/NIRCam, and JWST/NIRISS.
\begin{table*}[h!]
\centering
 \caption{All photometric bands used to correct the NIRSpec prism spectra for slit losses.
 \label{bands}
 }
\begin{tabular}{lcccccccc}
 \hline \hline
  \noalign{\smallskip}
  field & GDS & GDN & CEERS & ABELL2744  & RXJ2129 & MACS0647 \\
   \noalign{\smallskip}
 \hline
 \noalign{\smallskip}
 \noalign{\smallskip}
 \multicolumn{7}{c}{HST/ACS}   \\
  \hline
  \noalign{\smallskip}
 F435W &{\color{blue}  \ding{51}} & {\color{blue}  \ding{51}}  & {\color{blue}  \ding{51}} &  {\color{blue}  \ding{51}} & {\color{blue}  \ding{51}} & {\color{blue}  \ding{51}}  \\
 \noalign{\smallskip}
  F475W  & {\color{blue}  \ding{51}} & {\color{red}  \ding{55}}  & {\color{red}  \ding{55}}  & {\color{blue}  \ding{51}}  & {\color{blue}  \ding{51}} & {\color{blue}  \ding{51}}  \\
 \noalign{\smallskip}
  \noalign{\smallskip}
F555W  & {\color{red}  \ding{55}} & {\color{red}  \ding{55}}  &{\color{red}  \ding{55}}  & {\color{red}  \ding{55}} & {\color{blue}  \ding{51}} & {\color{blue}  \ding{51}}  \\
 \noalign{\smallskip}
F606W & {\color{blue}  \ding{51}} & {\color{blue}  \ding{51}}  & {\color{blue}  \ding{51}} &  {\color{blue}  \ding{51}}  & {\color{blue}  \ding{51}}& {\color{blue}  \ding{51}}  \\
 \noalign{\smallskip}
F625W  & {\color{red}  \ding{55}} & {\color{red}  \ding{55}}  &{\color{red}  \ding{55}}  & {\color{red}  \ding{55}} & {\color{blue}  \ding{51}} & {\color{blue}  \ding{51}} \\
 \noalign{\smallskip}
F775W & {\color{blue}  \ding{51}} & {\color{blue}  \ding{51}}  & {\color{red}  \ding{55}}  & {\color{blue}  \ding{51}}  & {\color{blue}  \ding{51}} & {\color{blue}  \ding{51}}  \\
 \noalign{\smallskip}
F814W & {\color{blue}  \ding{51}} & {\color{blue}  \ding{51}}  & {\color{blue}  \ding{51}} &  {\color{blue}  \ding{51}}  & {\color{blue}  \ding{51}} & {\color{blue}  \ding{51}}  \\
 \noalign{\smallskip}
F850LP & {\color{blue}  \ding{51}}  & {\color{blue}  \ding{51}} & {\color{red}  \ding{55}}  & {\color{red}  \ding{55}}  & {\color{blue}  \ding{51}}& {\color{blue}  \ding{51}}  \\
 \noalign{\smallskip}
  \hline
 \noalign{\smallskip}
 \multicolumn{7}{c}{HST/WFC3/UVIS}   \\
  \hline
   \noalign{\smallskip}
F275WU & {\color{red}  \ding{55}} & {\color{red}  \ding{55}} & {\color{red}  \ding{55}} &  {\color{red}  \ding{55}}  &{\color{red}  \ding{55}} & {\color{blue}  \ding{51}}  \\
 \noalign{\smallskip}
 F336WU & {\color{red}  \ding{55}} & {\color{red}  \ding{55}} & {\color{red}  \ding{55}} &  {\color{red}  \ding{55}}  &{\color{red}  \ding{55}} & {\color{blue}  \ding{51}}  \\
 \noalign{\smallskip}
 F350LPU & {\color{red}  \ding{55}} & {\color{red}  \ding{55}} & {\color{red}  \ding{55}} &  {\color{red}  \ding{55}}  &{\color{red}  \ding{55}} & {\color{blue}  \ding{51}}  \\
 \noalign{\smallskip}
 F390WU & {\color{red}  \ding{55}} & {\color{red}  \ding{55}} & {\color{red}  \ding{55}} &  {\color{red}  \ding{55}}  &{\color{red}  \ding{55}} & {\color{blue}  \ding{51}}  \\
 \noalign{\smallskip}
F606WU & {\color{blue}  \ding{51}}  & {\color{red}  \ding{55}} & {\color{blue}  \ding{51}} &  {\color{red}  \ding{55}}  &{\color{blue}  \ding{51}} & {\color{red}  \ding{55}}  \\
 \noalign{\smallskip}
F814WU & {\color{blue}  \ding{51}} & {\color{red}  \ding{55}} & {\color{red}  \ding{55}}  &  {\color{red}  \ding{55}}  & {\color{red}  \ding{55}} & {\color{red}  \ding{55}}  \\
 \noalign{\smallskip}
  \hline
   \noalign{\smallskip}
 \multicolumn{7}{c}{HST/WFC3}   \\
  \hline
   \noalign{\smallskip}
 F105W & {\color{blue}  \ding{51}}  & {\color{blue}  \ding{51}} & {\color{blue}  \ding{51}} &{\color{blue}  \ding{51}} & {\color{blue}  \ding{51}} & {\color{blue}  \ding{51}} \\
 \noalign{\smallskip}
 F110W & {\color{blue}  \ding{51}}  & {\color{blue}  \ding{51}} & {\color{red}  \ding{55}}  & {\color{red}  \ding{55}} & {\color{blue}  \ding{51}} & {\color{blue}  \ding{51}}  \\
 \noalign{\smallskip}
  F125W & {\color{blue}  \ding{51}}  & {\color{blue}  \ding{51}} & {\color{blue}  \ding{51}}  & {\color{blue}  \ding{51}} &{\color{blue}  \ding{51}} & {\color{blue}  \ding{51}}  \\
 \noalign{\smallskip}
  F140W & {\color{blue}  \ding{51}}  & {\color{blue}  \ding{51}} & {\color{blue}  \ding{51}}  & {\color{blue}  \ding{51}} & {\color{blue}  \ding{51}} & {\color{blue}  \ding{51}}  \\
 \noalign{\smallskip}
  F160W & {\color{blue}  \ding{51}}  & {\color{blue}  \ding{51}} & {\color{blue}  \ding{51}}  & {\color{blue}  \ding{51}}& {\color{blue}  \ding{51}} & {\color{blue}  \ding{51}}  \\
 \noalign{\smallskip}
  \hline
    \noalign{\smallskip}
 \multicolumn{7}{c}{JWST/NIRCam} \\
     \hline
 \noalign{\smallskip}
  F090W & {\color{blue}  \ding{51}}  & {\color{red}  \ding{55}} & {\color{red}  \ding{55}}  & {\color{blue}  \ding{51}} & {\color{red}  \ding{55}} & {\color{red}  \ding{55}} \\
 \noalign{\smallskip}
 F115W & {\color{blue}  \ding{51}}  & {\color{blue}  \ding{51}} & {\color{blue}  \ding{51}} & {\color{blue}  \ding{51}} & {\color{blue}  \ding{51}} & {\color{blue}  \ding{51}} \\
 \noalign{\smallskip}
 F150W & {\color{blue}  \ding{51}}  & {\color{blue}  \ding{51}}  & {\color{blue}  \ding{51}} & {\color{blue}  \ding{51}} & {\color{blue}  \ding{51}} & {\color{blue}  \ding{51}}  \\
 \noalign{\smallskip}
F182M & {\color{blue}  \ding{51}}  & {\color{blue}  \ding{51}}  & {\color{blue}  \ding{51}} & {\color{red}  \ding{55}} & {\color{red}  \ding{55}} & {\color{red}  \ding{55}} \\
 \noalign{\smallskip}
F200W & {\color{blue}  \ding{51}}  & {\color{red}  \ding{55}} & {\color{blue}  \ding{51}} & {\color{blue}  \ding{51}} &{\color{blue}  \ding{51}} & {\color{blue}  \ding{51}}  \\
 \noalign{\smallskip}
F210M & {\color{blue}  \ding{51}}  & {\color{blue}  \ding{51}} & {\color{blue}  \ding{51}} & {\color{red}  \ding{55}} & {\color{red}  \ding{55}} & {\color{red}  \ding{55}} \\
 \noalign{\smallskip}
F277W & {\color{blue}  \ding{51}}  & {\color{red}  \ding{55}} & {\color{blue}  \ding{51}} & {\color{blue}  \ding{51}} & {\color{blue}  \ding{51}} & {\color{blue}  \ding{51}} \\
 \noalign{\smallskip}
F335M & {\color{blue}  \ding{51}}  & {\color{red}  \ding{55}} & {\color{red}  \ding{55}}  & {\color{red}  \ding{55}} & {\color{red}  \ding{55}} & {\color{red}  \ding{55}} \\
 \noalign{\smallskip}
F356W &{\color{blue}  \ding{51}}  & {\color{blue}  \ding{51}}  & {\color{blue}  \ding{51}} & {\color{blue}  \ding{51}} & {\color{blue}  \ding{51}} & {\color{blue}  \ding{51}} \\
 \noalign{\smallskip}
F410M & {\color{blue}  \ding{51}}  & {\color{red}  \ding{55}}  & {\color{blue}  \ding{51}} & {\color{blue}  \ding{51}} & {\color{red}  \ding{55}} & {\color{red}  \ding{55}}  \\
 \noalign{\smallskip}
F430M & {\color{blue}  \ding{51}}  & {\color{red}  \ding{55}} & {\color{red}  \ding{55}}  & {\color{red}  \ding{55}} & {\color{red}  \ding{55}} & {\color{red}  \ding{55}}\\
 \noalign{\smallskip}
F444W & {\color{blue}  \ding{51}}  & {\color{blue}  \ding{51}}  &  {\color{blue}  \ding{51}} & {\color{blue}  \ding{51}} & {\color{blue}  \ding{51}} & {\color{blue}  \ding{51}} \\
 \noalign{\smallskip}
F460M & {\color{blue}  \ding{51}}  & {\color{red}  \ding{55}}  & {\color{red}  \ding{55}}  & {\color{red}  \ding{55}} & {\color{red}  \ding{55}} & {\color{red}  \ding{55}}  \\
 \noalign{\smallskip}
F480M & {\color{blue}  \ding{51}}  & {\color{red}  \ding{55}} & {\color{red}  \ding{55}}  & {\color{red}  \ding{55}} & {\color{red}  \ding{55}} & {\color{blue}  \ding{51}} \\
 \noalign{\smallskip}
 \hline
    \noalign{\smallskip}
 \multicolumn{7}{c}{JWST/NIRISS} \\
     \hline
 \noalign{\smallskip}
F115WN & {\color{blue}  \ding{51}}  & {\color{red}  \ding{55}}  & {\color{red}  \ding{55}}  & {\color{blue}  \ding{51}} & {\color{red}  \ding{55}} & {\color{red}  \ding{55}} \\
 \noalign{\smallskip}
F150WN & {\color{blue}  \ding{51}}  & {\color{red}  \ding{55}}  & {\color{red}  \ding{55}} & {\color{blue}  \ding{51}} & {\color{red}  \ding{55}} & {\color{red}  \ding{55}} \\
 \noalign{\smallskip}
F200WN & {\color{blue}  \ding{51}}  & {\color{red}  \ding{55}}  & {\color{red}  \ding{55}} & {\color{blue}  \ding{51}}  & {\color{red}  \ding{55}} & {\color{red}  \ding{55}}  \\
 \noalign{\smallskip}
 \noalign{\smallskip}
  \hline
\end{tabular}
\end{table*}

\FloatBarrier

\section{Dust-age-metallicity degeneracy}
\label{degeneracy}

The dust–age–metallicity degeneracy is a well-known limitation of any SED fitting method, particularly in the absence of far-infrared (FIR) observations, as is the case for our dataset (see the Methodology section of \citealp{2024NatAs.tmp...20M}). However, even when rest-frame FIR data are available (e.g., from ALMA), they do not necessarily trace the same regions of a galaxy traced by the rest-frame optical emission. Significant spatial offsets between stellar and dust-emitting regions have been found at high redshift (e.g., \citealp{2018MNRAS.477..552B, 2018MNRAS.478.1170C, 2024arXiv240218543F, 2024ApJ...977L..36H}), complicating a direct interpretation of the total energy budget.

To assess the influence of these degeneracies on our main results, we first tested the robustness of the key trends presented in this work — namely, the $S$–$A_V$, $B$–$A_V$, and $S$–${\langle a \rangle}_*^{\rm{m}}$ relations — to variations in the two remaining degenerate parameters.
As shown in Figure \ref{SB_Av_age_const} (Appendix \ref{App_AV}), the $S$–$A_V$ and $B$–$A_V$ trends are largely independent of stellar age, while the $S$–${\langle a \rangle}_*^{\rm{m}}$ relation is not  affected by $A_V$ (Figure \ref{S_reds_age}, Appendix \ref{App_age}). In addition, we verified that these trends remain robust against variations in metallicity ($Z$).

To more directly quantify the impact of SFH-related degeneracies, particularly the age-attenuation degeneracy, we re-ran the SED fitting for the full galaxy sample using a parametric constant SFH model. This contrasts with our fiducial model Sect. \ref{method_SED}; \citealp{2023A&A...679A..12M, 2024NatAs.tmp...20M}), which adopts a flexible non-parametric (step-function) SFH (a comparison often adopted in the literature, e.g., \citealp{2019ApJ...876....3L, 2022MNRAS.516..975T, 2023MNRAS.519.5859W}). This test allows us to evaluate how sensitive the derived parameters are to the assumed SFH shape, thereby directly highlighting the impact of the age-attenuation degeneracy.

We found that the key trends identified in this work: $S$–$A_V$, $B$–$A_V$, and $S$–${\langle a \rangle}_*^{\rm{m}}$, as well as those reported in \cite{2024NatAs.tmp...20M} ($S-z$ and $B-z$), remain largely present when using the constant SFH model, although the correlations are generally somewhat weaker, compared to our fiducial run. Most trends retain moderate Pearson correlation coefficients ($|r|\geq 0.3$), except for $B-z$, which becomes weaker ($r \sim -0.19$), consistent with its already lower significance in the fiducial analysis ($r \sim -0.28$; Fig. \ref{PCA}). 

We also repeated the RFR analysis using the outputs from the constant SFH run and found results broadly consistent with those from the fiducial model (Fig. \ref{RFR}). The most important predictors of $S$ remain redshift ($\sim 0.21$), $A_V$ ($\sim 0.31$), and stellar age (${\langle a \rangle}_*^{\rm{m}}$; $\sim 0.16$), while $A_V$ remains the dominant predictor of $B$ ($\sim 0.27$).

\begin{figure*}[h!]
\centering
\includegraphics[width=0.49\hsize]{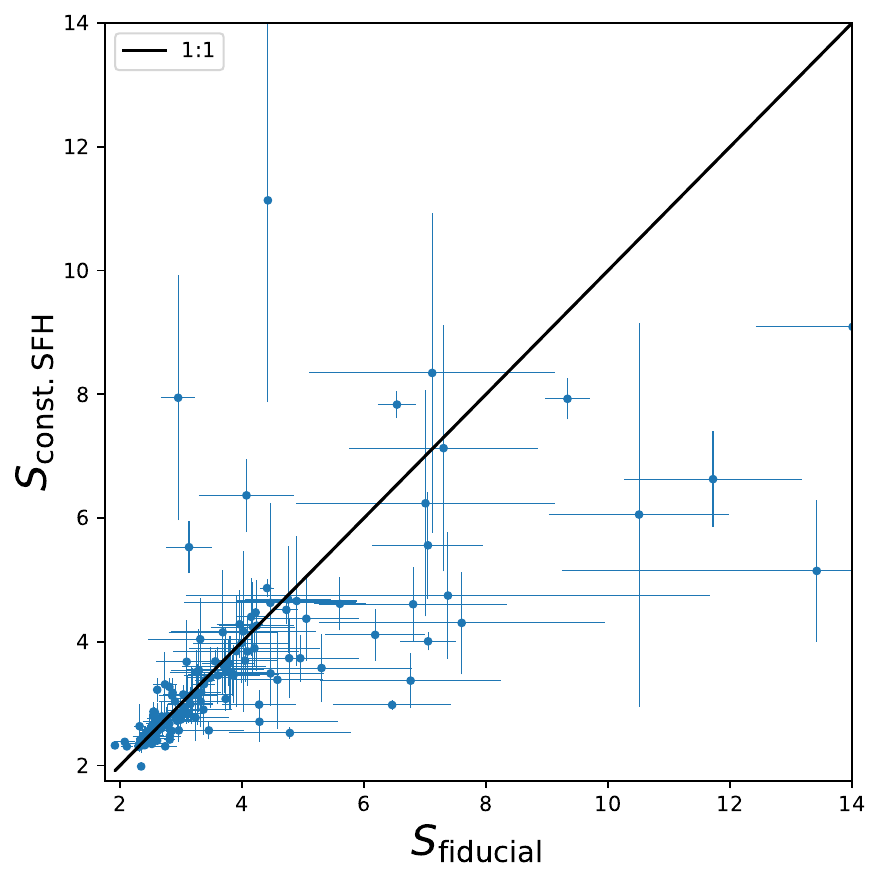}
\includegraphics[width=0.49\hsize]{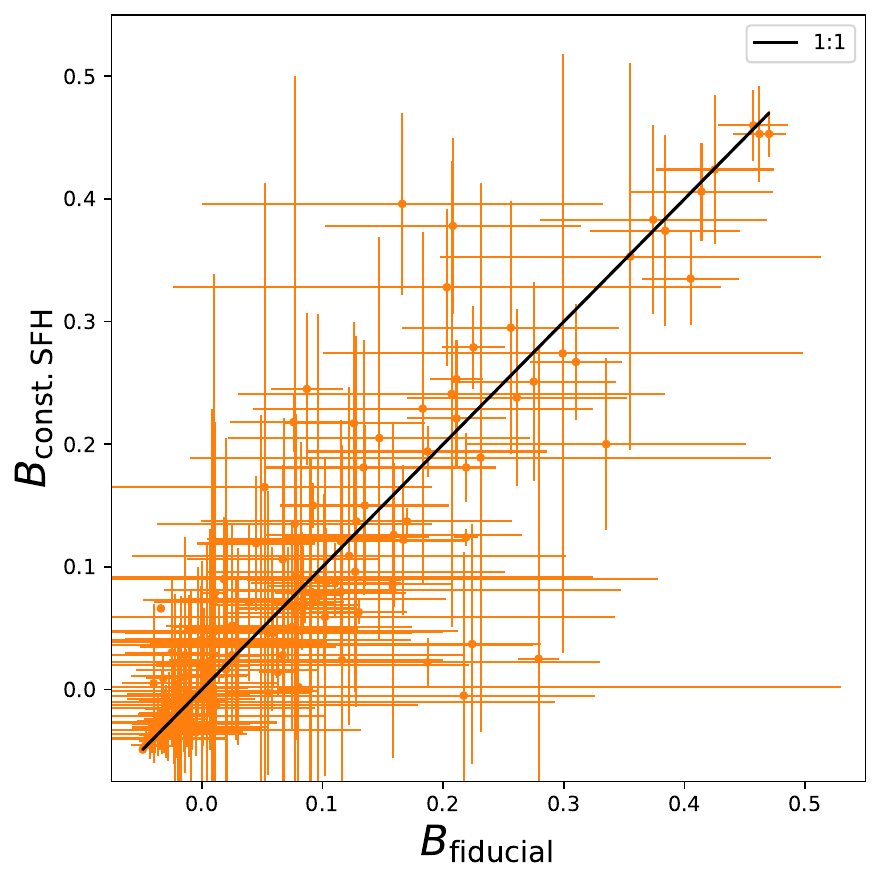}
\caption{Dust attenuation curve parameters $S$ and $B$ derived from SED fitting using two different star formation histories: a constant SFH and the fiducial flexible (non-parametric) SFH model (\citealp{2019ApJ...876....3L}). The left panel shows the slope parameter $S$ (blue), and the right panel shows the UV bump strength $B$ (orange). The solid black line indicates the 1:1 relation.  
\label{SS_BB}
}
\end{figure*}

To further assess the robustness of the attenuation curve parameters themselves, we compared the slope $S$ and UV bump strength $B$ between the two SFH models for all 173 matched galaxies. We find that both parameters are overall consistent, with $S^{\rm{const SFH}} \in S^{\rm{fiducial}} \pm S_{\rm{err}}^{\rm{fiducial}}$ for $62\%$ of sources and $B^{\rm{const SFH}} \in B^{\rm{fiducial}} \pm B_{\rm{err}}^{\rm{fiducial}}$ for $84\%$ of galaxies (Fig. \ref{SS_BB}). These consistency levels indicate that both $S$ and $B$ are relatively stable across different SFH assumptions. This aligns with our earlier findings that the dust attenuation curve shapes are largely independent of the adopted SFH model, based on a smaller galaxy sample (\citealp{2023A&A...679A..12M}).

The most significant discrepancy between the two fitting runs is seen in the ${\langle a \rangle}_*^{\rm{m}}$ parameter. Under the constant SFH model, stellar age estimates are strongly biased toward younger values due to the well-documented outshining effect (e.g., \citealp{2019ApJ...876....3L, 2022MNRAS.516..975T, 2023MNRAS.519.5859W}). 

In summary, our analysis shows that the key physical trends and the inferred shape of the dust attenuation curve (as parametrized by the $S$ and $B$ parameters) are largely robust. While stellar age estimates are sensitive to SFH assumptions, the primary conclusions of our study remain unchanged.

\section{Slope ($S$) vs. sSFR}  \label{S_sSFR_trend}

\begin{figure}[h!]
\centering
\includegraphics[width=0.8\hsize]{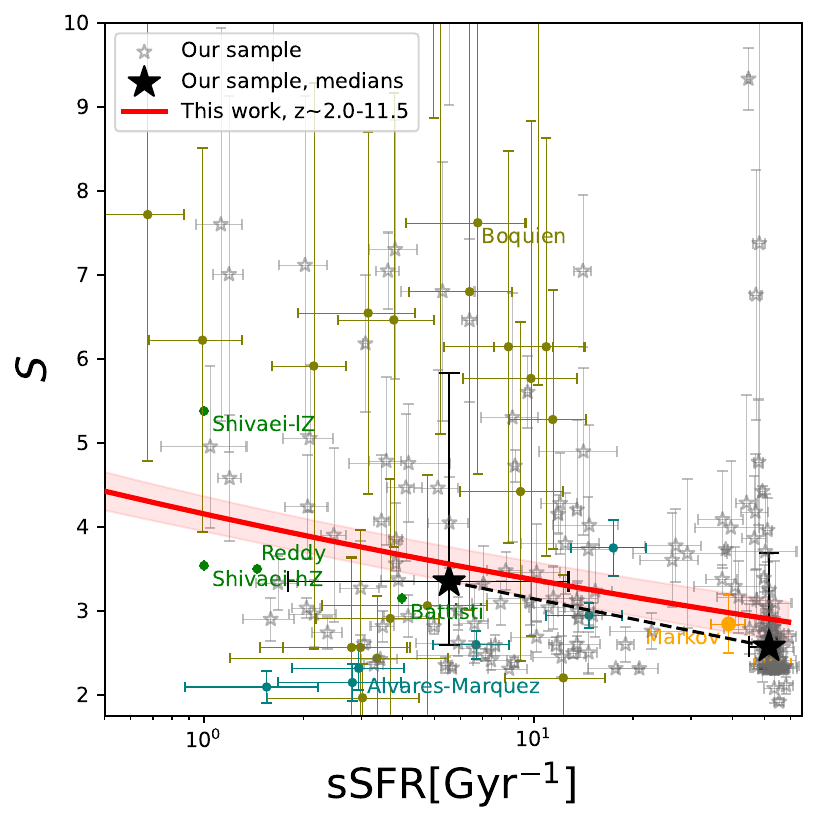}
\caption{UV-optical slope (S) as a function of the sSFR.   
Gray stars represent individual $S$ measurements from our sample, with $1\sigma$ uncertainties estimated via bootstrapping. Black stars show the median $S$ values for galaxies grouped by sSFR, with error bars indicating the $1\sigma$ dispersion within each bin. The solid red line shows the best-fit relation for the full sample using a power-law model (Eq. \ref{eq1}), and the shaded region denotes the associated uncertainty. Literature results for intermediate-$z$ sources or stacks are shown in green (\citealp{2015ApJ...806..259R,  2020ApJ...899..117S, 2022MNRAS.513.4431B})  and teal (\citealp{2019A&A...630A.153A}), while high-$z$ objects (\citealp{2022A&A...663A..50B, 2023A&A...679A..12M}) are displayed as olive and orange symbols, respectively. Error bars represent their $1\sigma$ uncertainties. 
\label{sSFR_S}
}
\end{figure}

Pearson's correlation analysis reveals a moderate negative correlation between $S$ and sSFR ($r = -0.38$), indicating that the slope $S$ tends to flatten with increasing sSFR. However, the RFR analysis suggests that sSFR is not a significant predictor of $S$ when other galaxy properties, such as redshift and stellar age, are included. The $S-\rm{sSFR}$ relation is shown in Fig. \ref{sSFR_S}. We fit the trend with a power law model (see Eq. \ref{eq1} and the corresponding best-fit coefficients are given in Table \ref{params}.

\FloatBarrier

\section{Best-fit parameters of the observed trends} \label{App_par}
Table \ref{params} summarizes the best-fit parameters for the observed trends between $S$ and $A_V$, and $B$ and $A_V$  (Sect. \ref{curve_Av_trend}), derived using Eq. \ref{eq1} and Eq. \ref{eqB}, respectively. We also report the best-fit parameters for $S-{\langle a \rangle}_*^{\rm{m}}$ (Sect. \ref{S_age_trend}), $S-{\rm{sSFR}}$, and ${\langle a \rangle}_*^{\rm{m}}-{\rm{z}}$ trends (Sect. \ref{global_trends}) all derived using Eq. \ref{eq1}. 

\begin{table*}[h!]
\centering
 \caption{Best-fit parameters of the various correlations found in our analysis.
 \label{params}
 }
\begin{tabular}{lcccccccc}
 \hline \hline
  \noalign{\smallskip}
  parameter & full sample & ... & ... & ... & ... \\
   \noalign{\smallskip}
$z$ & $2-12$ & $2-3.3$ &  $3.3-4.5$ & $4.5-5.7$ & $5.7-11.5$\\
\noalign{\smallskip}
 \hline
 \noalign{\smallskip}
 \multicolumn{6}{c}{$S-A_V$ correlation} \\
  \hline
 \noalign{\smallskip}
$k_s$ & $-0.32 \pm 0.04$  & $-0.59 \pm 0.11$ & $-0.17 \pm 0.06$ & $-0.19 \pm 0.05$ & $-0.10 \pm 0.03$  \\
\noalign{\smallskip}
 $n_s$ & $0.39 \pm 0.02$  &  $0.36 \pm 0.06$ & $0.49 \pm 0.04$ & $0.38 \pm 0.02$ & $0.37 \pm 0.01$ \\
 \noalign{\smallskip}
  \hline
 \noalign{\smallskip}
 \multicolumn{6}{c}{$B-A_V$ correlation}   \\
  \hline
 \noalign{\smallskip}
$k_b$ & $-0.13 \pm 0.03$ & $-0.29 \pm 0.10$ & $-0.06 \pm 0.05$ & $-0.14 \pm 0.05$ & $-0.10 \pm 0.05$\\
 \noalign{\smallskip}
$n_b$  & $0.14 \pm 0.02$ & $0.24 \pm 0.04$ & $0.13 \pm 0.03$ & $0.12 \pm 0.03$ & $0.09 \pm 0.03$  \\ %
 \noalign{\smallskip}
   \hline
    \noalign{\smallskip}
 \multicolumn{6}{c}{$S-{\langle a \rangle}_*^{\rm{m}}$ correlation} \\
     \hline
 \noalign{\smallskip}
$k_s$ & $0.12 \pm 0.02$  & $0.12 \pm 0.06$ & $0.05 \pm 0.03$ & $0.05 \pm 0.02$ & $0.04 \pm 0.02$ \\
\noalign{\smallskip}
 $n_s$ & $0.67 \pm 0.02$  &  $0.76 \pm 0.06$ & $0.62 \pm 0.03$ & $0.52 \pm 0.03$ & $0.48 \pm 0.03$ \\
 \noalign{\smallskip}
  \hline
    \noalign{\smallskip}
 \multicolumn{6}{c}{$S-{\rm{sSFR}}$ correlation} \\
     \hline
 \noalign{\smallskip}
$k_s$ & $-0.24 \pm 0.04$  & ... & ... & ... & ...  \\
\noalign{\smallskip}
 $n_s$ & $0.73 \pm 0.04$  & ... & ... & ... & ... \\
 \noalign{\smallskip}
  \hline
\noalign{\smallskip}
 \multicolumn{6}{c}{${\langle a \rangle}_*^{\rm{m}}-{\rm{z}}$ correlation} \\
     \hline
 \noalign{\smallskip}
$k_s$ & $-2.19 \pm 0.27$  & ... & ... & ... & ...  \\
\noalign{\smallskip}
 $n_s$ & $3.11 \pm 0.18$  & ... & ... & ... & ... \\
 \noalign{\smallskip}
  \hline
\end{tabular}
\end{table*}

\FloatBarrier

\section{Testing the consistency of $S-A_V$ and $B-A_V$ correlations}\label{App_AV}

\begin{figure*}[h!]
\centering
\includegraphics[width=0.49\hsize]{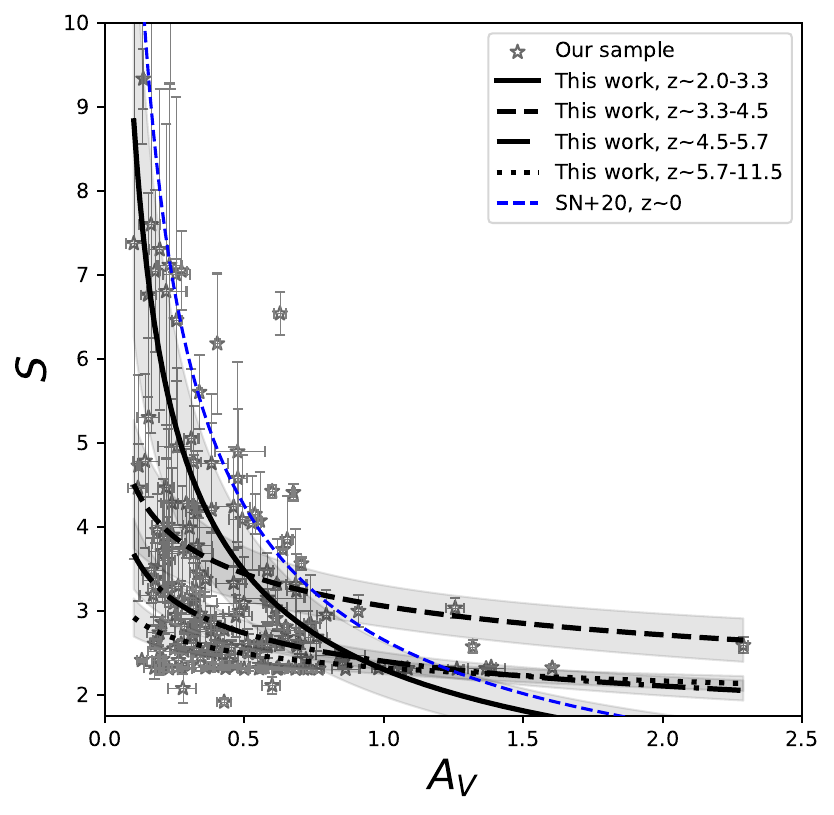}
\includegraphics[width=0.49\hsize]{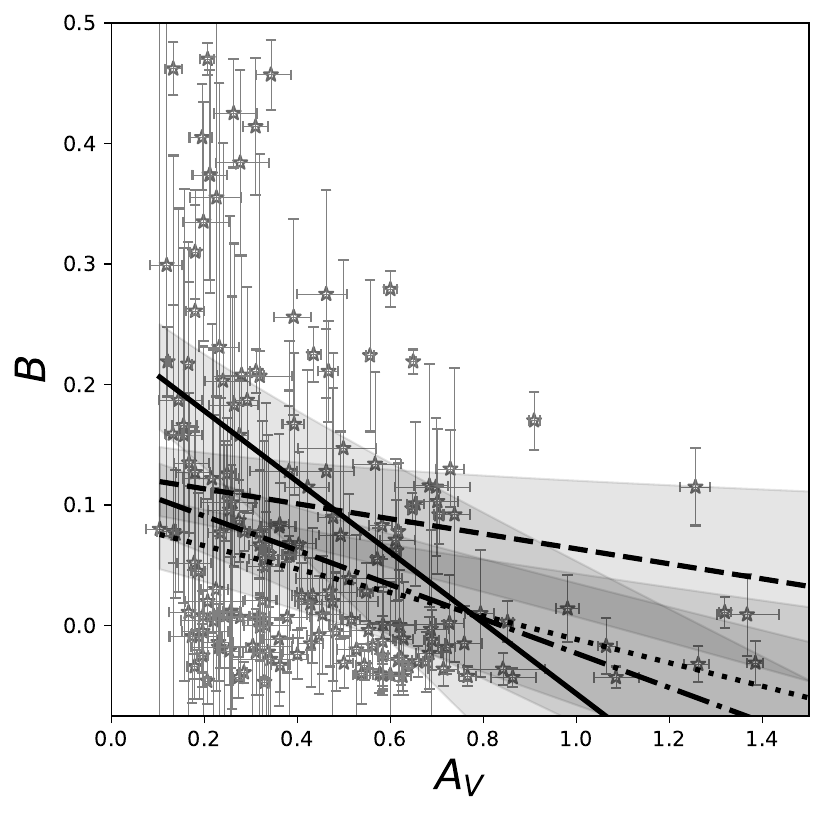}
\caption{Dust attenuation parameters as a function of $V$-band attenuation, $A_V$. Our sample is represented as gray stars. 
The left panel shows the UV-optical slope (S), while the right panel displays the UV bump strength (B) as a function of $A_V$.
Black lines represent the best-fit correlations and their $1\sigma$ uncertainties for the subsets grouped by redshift, with shaded regions indicating the uncertainty ranges. The $S-A_V$ correlation from \cite{2020ARA&A..58..529S} for a nearby ($z \sim 0$) sample is shown as a blue dashed line in the left panel.
\label{SB_Av_z_const}
}
\end{figure*}

\begin{figure*}[h!]
\centering
\includegraphics[width=0.49\hsize]{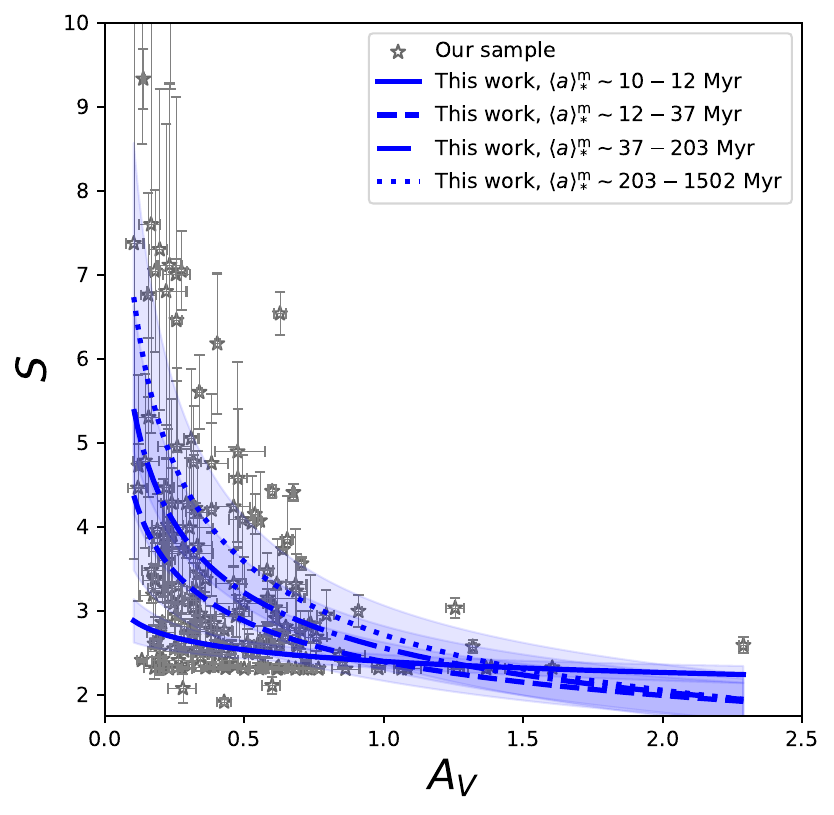}
\includegraphics[width=0.49\hsize]{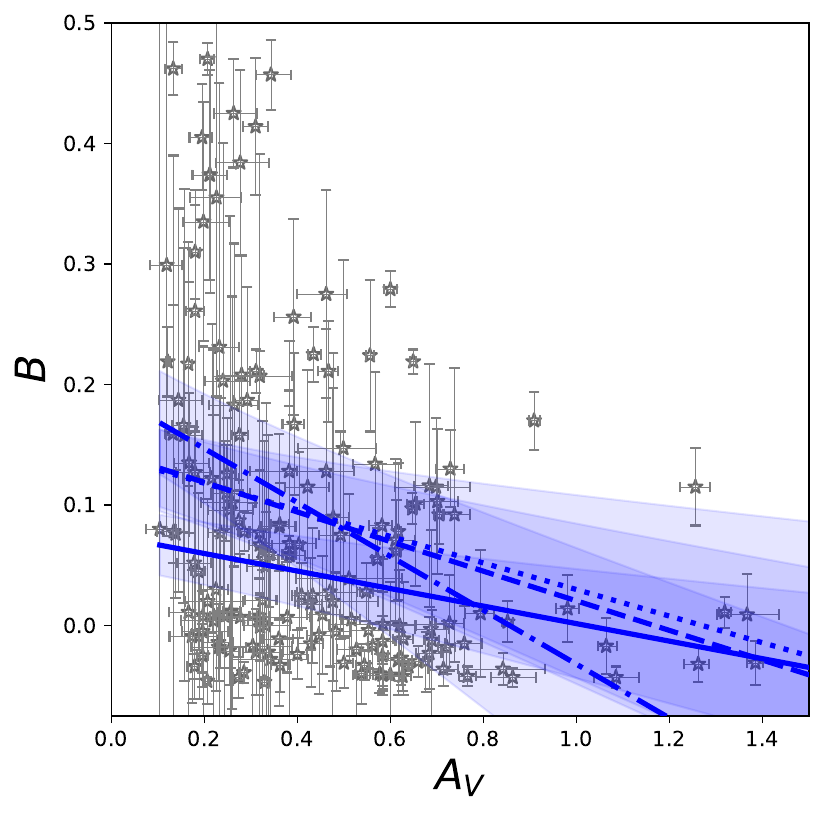}
\caption{Dust attenuation parameters as a function of $V$-band attenuation, $A_V$. Our sample is represented as gray stars. 
The left panel shows the UV-optical slope (S), while the right panel displays the UV bump strength (B) as a function of $A_V$.
Blue lines show the best-fit correlations and their $1sigma$ uncertainties for the subsets grouped by stellar ages, with shaded regions indicating the uncertainty ranges. 
\label{SB_Av_age_const}
}
\end{figure*}

\begin{figure*}[h!]
\centering
\includegraphics[width=0.49\hsize]{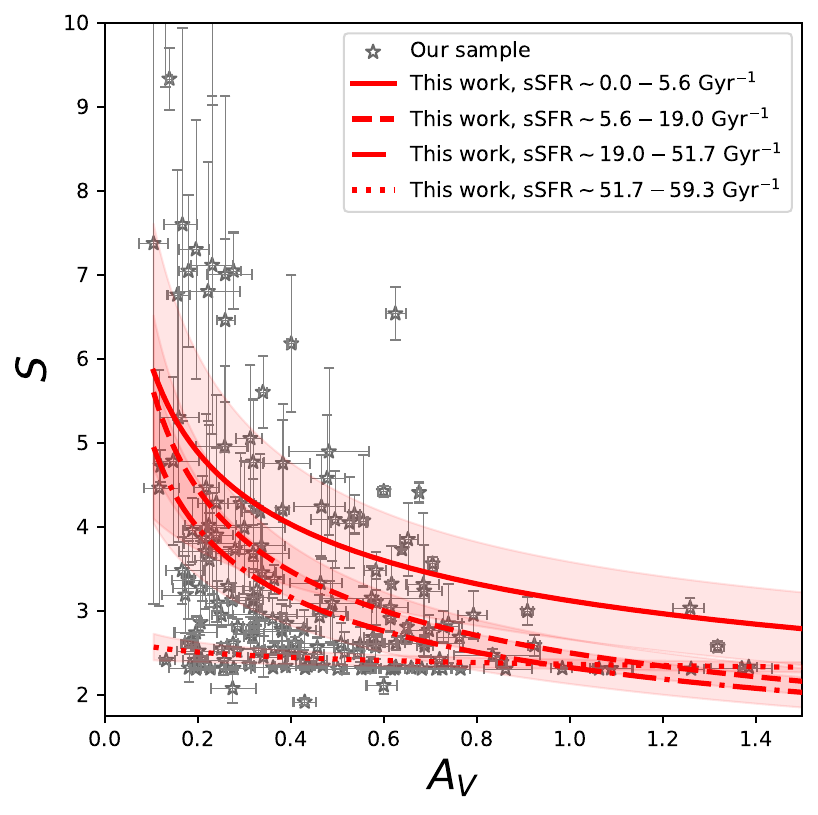}
\includegraphics[width=0.49\hsize]{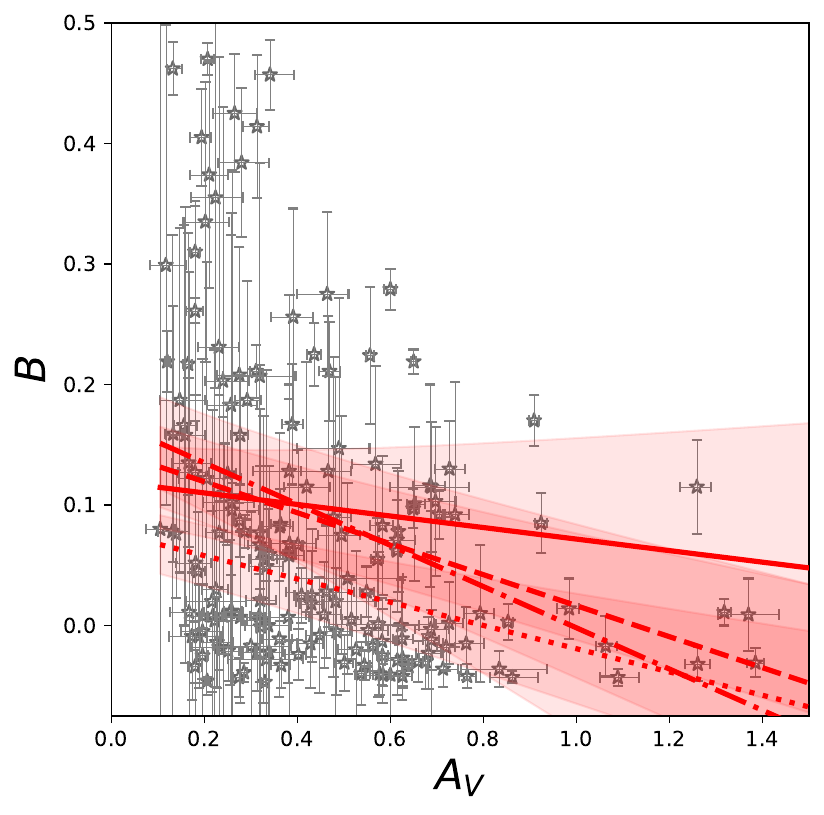}
\caption{Dust attenuation parameters as a function of $V$-band attenuation, $A_V$. Our sample is represented as gray stars. 
The left panel shows the UV-optical slope (S), while the right panel displays the UV bump strength (B) as a function of $A_V$.
Red lines show the best-fit correlations and their $1sigma$ uncertainties for the subsets grouped by sSFR, with shaded regions indicating the uncertainty ranges. 
\label{SB_Av_sSFR_const}
}
\end{figure*}

To assess whether the $S-A_V$ and $B-A_V$ trends are generally independent of the remaining key drivers of variations in $S$ and $B$ (i.e., redshift, ${\langle a \rangle}_*^{\rm{m}}$, and sSFR) we divided our full sample ($z \sim 2-11.5$), into four subsets,  ensuring approximately equal numbers of sources per bin for each property. 
In Figs. \ref{SB_Av_z_const}, \ref{SB_Av_age_const}, and \ref{SB_Av_sSFR_const} we illustrate that the $S-A_V$ and $B-A_V$ trends persist across the redshift, mass-weighted stellar age (${\langle a \rangle}_*^{\rm{m}}$) and sSFR subsets, respectively. These trends are primarily driven by RT effects related to the amount and spatial distribution of dust relative to stars.

\FloatBarrier

\section{Testing the consistency of $S-{\langle a \rangle}_*^{\rm{m}}$ correlation} \label{App_age}

\begin{figure*}[h!]
\centering
\includegraphics[width=0.33\hsize]{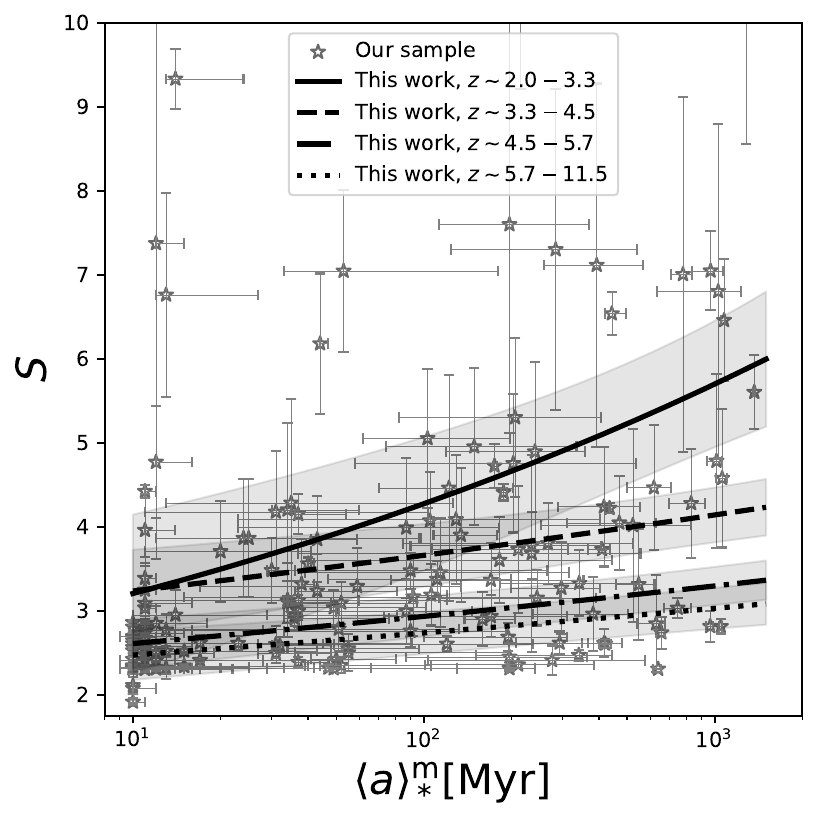}
\includegraphics[width=0.33\hsize]{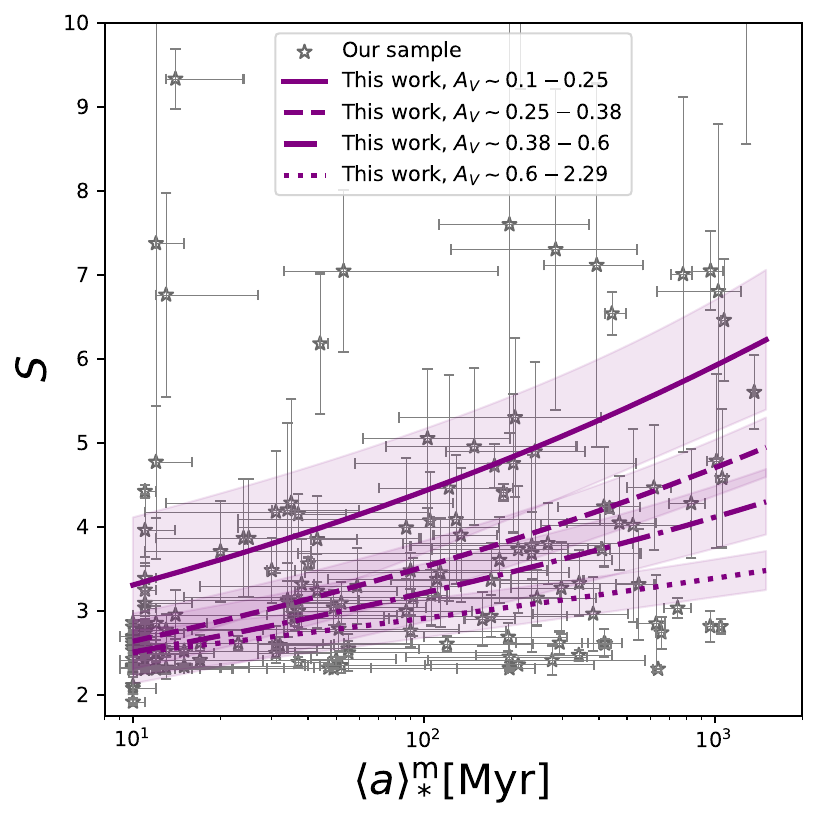}
\includegraphics[width=0.33\hsize]{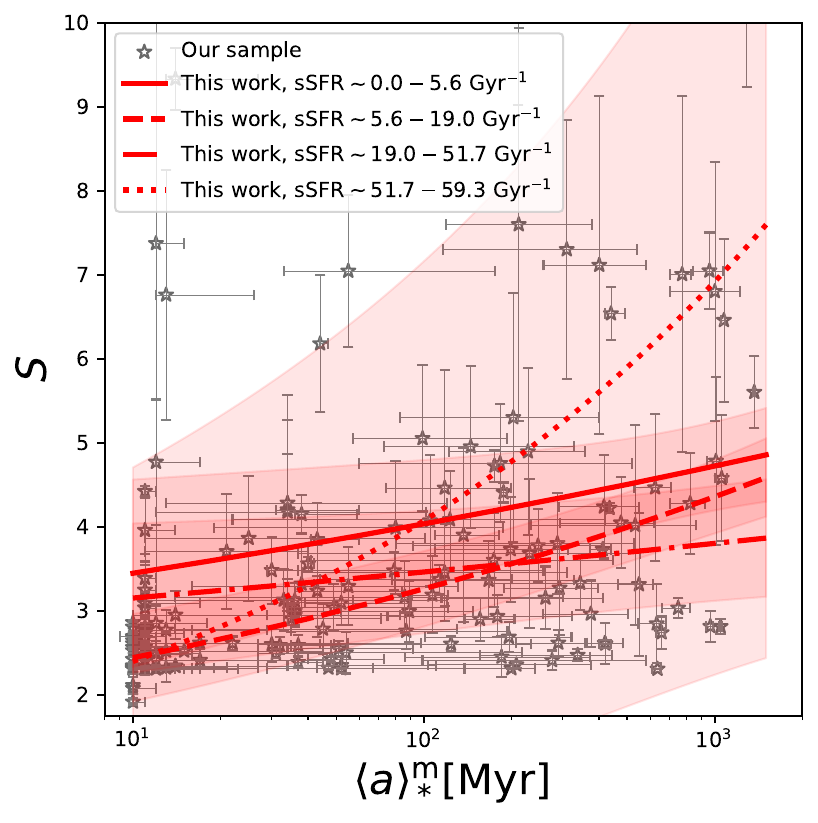}
\caption{UV-optical slope (S) as a function of the mass-weighted stellar age (${\langle a \rangle}_*^{\rm{m}}$).   
The black, purple, and red lines represent the best-fit correlations and their $1\sigma$ uncertainties for the subsets grouped by $z$, $A_V$, and sSFR (left, middle, and right panel, respectively). Shaded regions indicate the uncertainty ranges.
\label{S_age_AvsSFR}
}
\end{figure*}

\begin{figure*}[h!]
\centering
\includegraphics[width=0.49\hsize]{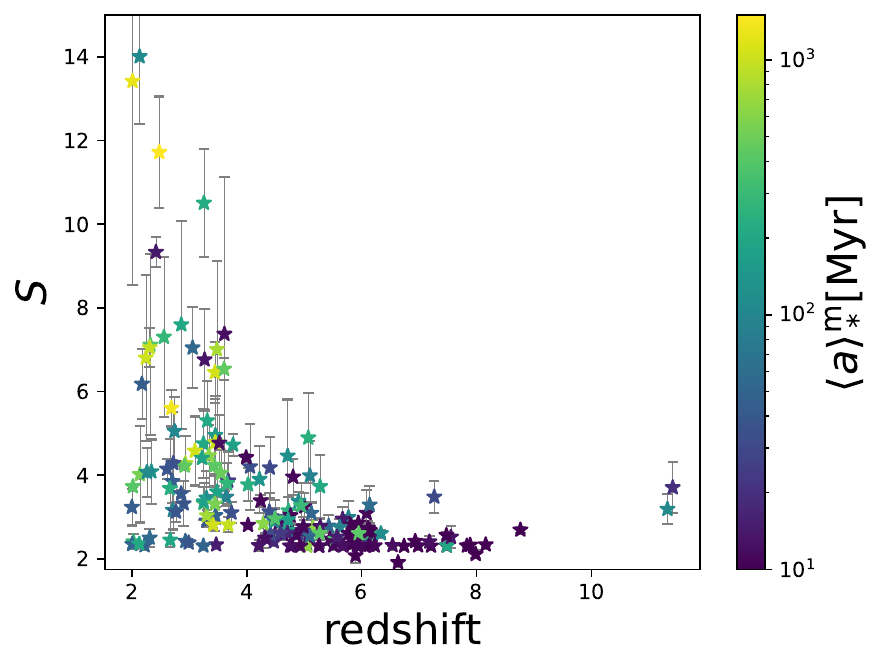}
\includegraphics[width=0.49\hsize]{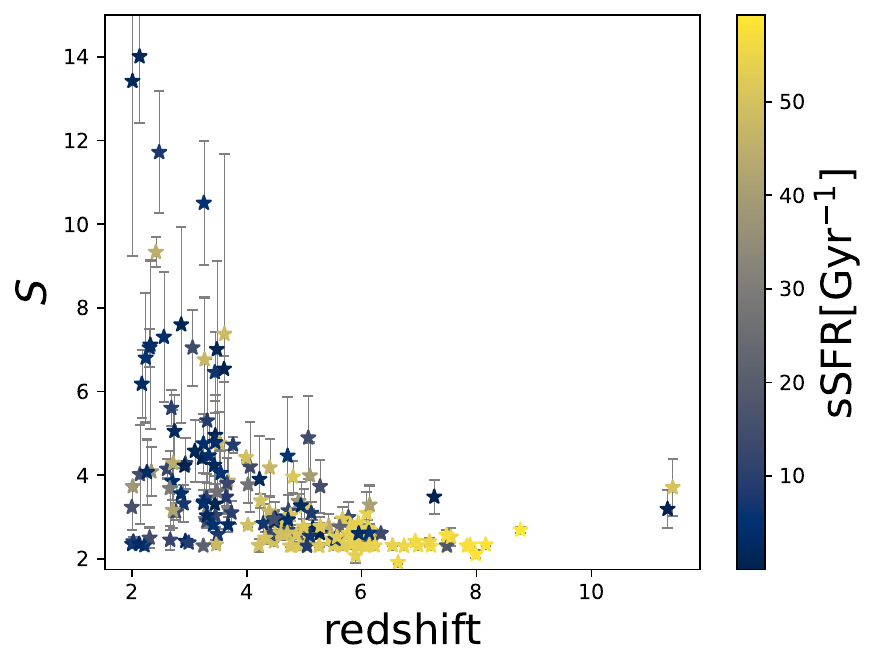}
\caption{UV-optical slope ($S$) vs. the redshift. Our sample is represented as stars, color-coded by the mass-weighted stellar age (${\langle a \rangle}_*^{\rm{m}}$; left panel), and sSFR (right panel).
\label{S_reds_age}
}
\end{figure*}

Similarly to Appendix \ref{App_AV}, to investigate the persistency of the $S-{\langle a \rangle}_*^{\rm{m}}$ correlation on other key drivers of $S$ variations (i.e., $z$, $A_V$, and sSFR) we divide the full galaxy sample into bins based on each of these parameters.
The left panel of Fig. \ref{S_age_AvsSFR} illustrates $S-{\langle a \rangle}_*^{\rm{m}}$ trends for subsets grouped by redshift. Fig. \ref{S_age_AvsSFR} shows that the $S-{\langle a \rangle}_*^{\rm{m}}$ trends become less pronounced when restricting the analysis to a relatively narrow redshift range, with a slope significance of  ($k_s^{{\langle a \rangle}}/(k_s^{{\langle a \rangle}})_{err} \sim 2$; Table \ref{params}),  suggesting that the redshift evolution of the intrinsic dust properties likely contributes to the observed $S-{\langle a \rangle}_*^{\rm{m}}$ trends. 

Next, the middle panel of Fig. \ref{S_age_AvsSFR} shows that the $S-{\langle a \rangle}_*^{\rm{m}}$ trend remains robust regardless of $A_V$ - a proxy for RT effects, indicating that the $S-{\langle a \rangle}_*^{\rm{m}}$ trends are not primarily driven by dust–star geometry. 

Finally, the right panel of Fig. \ref{S_age_AvsSFR}) shows that no statistically significant $S-{\langle a \rangle}_*^{\rm{m}}$ trends are observed when controlling for sSFR, except in the second sSFR bin (which spans a relatively wide range,  $\rm{sSFR} \sim 5.6-19.0 \ {Gyr}^{-1}$). This is expected as ${\langle a \rangle}_*^{\rm{m}}$ and sSFR are strongly correlated, indicating the origin of the $S-{\langle a \rangle}_*^{\rm{m}}$ trend is likely linked to variations in the strength of the radiation field.

\end{appendix}
\end{document}